\documentclass[aps,prb,superscriptaddress,showpacs]{revtex4}

\usepackage{graphics}
\usepackage{bm}

\usepackage{amsmath}    
\usepackage{amssymb}    
\usepackage{graphicx}   
\usepackage{subfigure}
\begin{document}

\title {Static Properties of Small Josephson Tunnel Junctions in an Oblique Magnetic Field}
\thanks{To be published in Phys. Rev. B}

\author{R. Monaco}
\affiliation{Istituto di Cibernetica del CNR, 80078, Pozzuoli, Italy
and Unit$\grave{\rm a}$ INFM $­$ Dipartimento di Fisica, Universit$\grave{\rm a}$ di Salerno, 84081 Baronissi, Italy}\email{roberto@sa.infn.it}
\author{M. Aaroe}
\affiliation{DTU Physics, B309, Technical University of
Denmark, DK-2800 Lyngby, Denmark}
\author{J. Mygind}
\affiliation{DTU Physics, B309, Technical University of
Denmark, DK-2800 Lyngby, Denmark}
\author{V.P. Koshelets}
\affiliation{Kotel'nikov Institute of Radio Engineering and Electronics, Russian
Academy of Science, Mokhovaya 11, B7, 125009, Moscow, Russia}

\date{\today}

\begin{abstract}
We have carried out a detailed experimental investigation of the
static properties of planar Josephson tunnel junctions in presence
of a uniform external magnetic field applied in an arbitrary
orientation with respect to the barrier plane. We considered annular
junctions, as well as rectangular junctions (having both overlap and
cross-type geometries) with different barrier aspect ratios. It is shown
how most of the experimental findings in an oblique field can be
reproduced invoking the superposition principle to combine the
classical  behavior of electrically small junctions in an in-plane field together
with the small junction behavior in a transverse field that we
recently published [R. Monaco, et al,, {\it J. Appl. Phys.} {\bf 104}, 023906 (2008)]. We explore the implications of these results in supposing systematic
errors in previous experiments and in proposing new possible
applications. We show that the presence of a transverse field may have
important consequences, which could be either voluntarily exploited
in applications or present an unwanted perturbation.
\end{abstract}

\pacs{74.50.+r}
\maketitle

\section{Introduction\label{sec:Intro}}

\begin{figure}[ht]
        \centering
                \includegraphics[width=9cm]{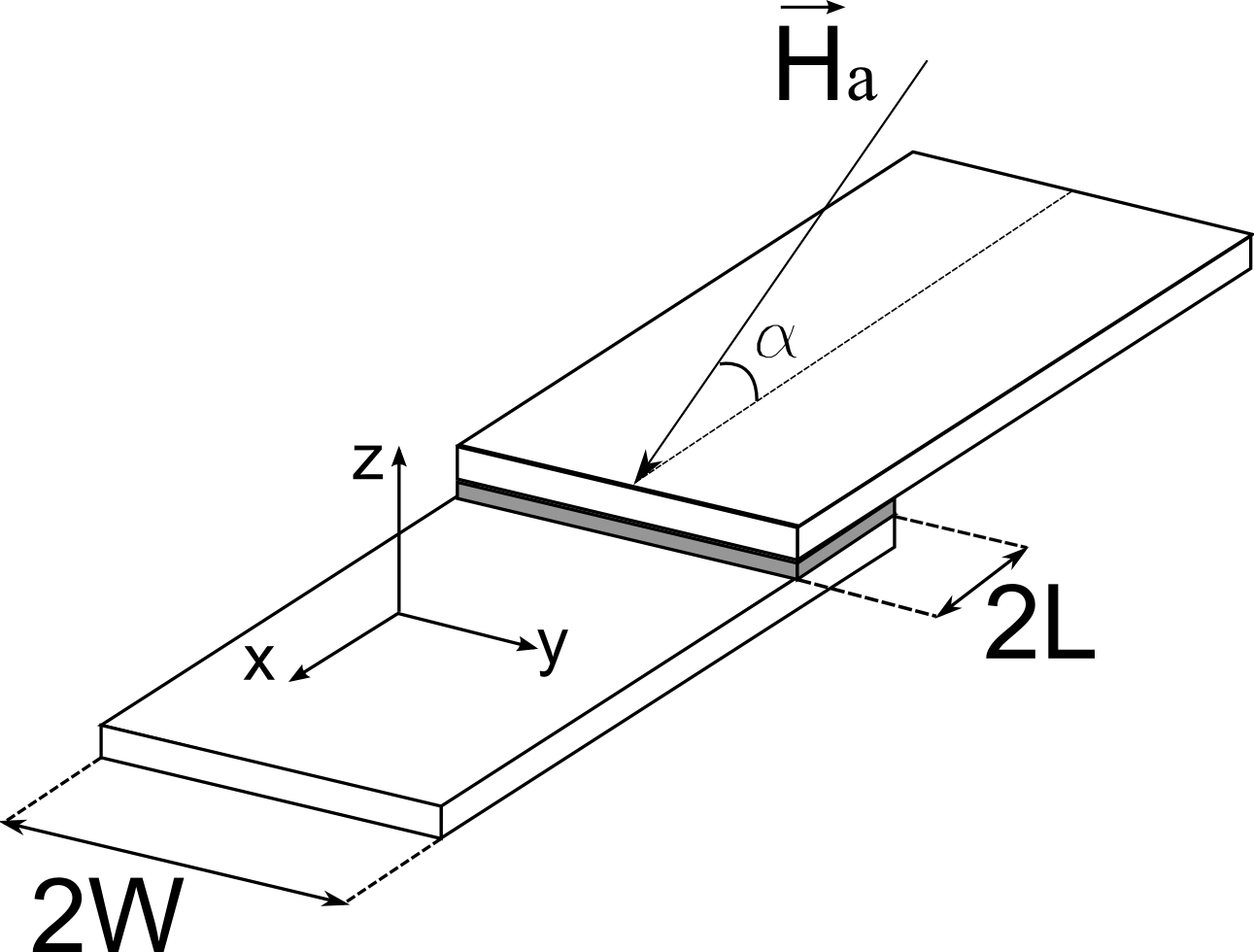}
        \caption{Sketch of a rectangular planar Josephson tunnel junction. The tunnel barrier lies in the $z=0$ plan. An oblique field ${\bf H_a}$ is applied in the $x$-$z$ plane and forms an angle $\alpha$ with respect to the $x$-direction.}
        \label{Geometry}
\end{figure}

One of the earliest experiments involving Josephson junctions and
magnetic fields has been the measurement of the magnetic diffraction
pattern\cite{Rowell}, i.e., the dependence of the junction critical
current $I_c$ on the amplitude of an externally applied magnetic
field $\bf H_a$. Traditionally, since the discovery of the Josephson
effect in 1962, the magnetic diffraction pattern $I_c(H_a)$ of
planar Josephson tunnel junctions (JTJs) has been recorded with the
magnetic induction field applied in the junction plane to avoid the
huge computational complications of taking demagnetization effects
into account, when a transverse magnetic component is present. A
number of important results have been derived from experiments under
these assumptions - a prominent example being the determination of
the London penetration depth\cite{Broom} $\lambda_L$ from which one
derives the Josephson penetration depth\cite{Jos1964} $\lambda_J$
which sets the JTJ electric length scale. Nowadays, every textbook
on the Josephson effect deserves at least one chapter to the
magnetic diffraction phenomena. The simplest case is that sketched in Fig.\ref{Geometry} of a
rectangular JTJ placed in a uniform and constant external magnetic field parallel to
one of the barrier edges. Let us choose the coordinate system such
that the tunnel barrier lies in the $z=0$ plane and let $2L$
and $2W$ be the junction dimensions along the $x$ and $y$-directions,
respectively. Finally, let us assume that the JTJ is electrically
small, meaning that its dimensions are both smaller than the
Josephson penetration depth ($2L,2W<\lambda_J$)(absence of
self-fields) and that its Josephson current density $J_J$ is constant 
over the barrier area.

If the externally applied field ${\bf H_a}$ is along the $x$
direction, ${\bf H_a} = {\rm H_x} {\bf \hat x}$, then the magnetic
field ${\bf H}$ inside the junction is constant and equal to the
external value, i.e., ${\bf H} \equiv ( H_x, 0, 0)$. By
integrating the Josephson equation\cite{Jos1964} relating the
Josephson phase $\phi$ to the magnetic induction field in the
barrier $\bf H$:

\begin{equation}
\label{gra}
{\bf \nabla} \phi(x,y) = \frac{2\pi d_e \mu _0}{\Phi_0} {\bf H}(x,y)\times {\bf \hat z},
\end{equation}

\noindent in which $d_e$ is the junction magnetic thickness, $\mu_0$ is the vacuum permeability and
$\Phi_0=h/2e$ is the magnetic flux quantum, we readily obtain the
spatial dependence of the Josephson phase:

\begin{equation}
\label{py} \phi= \kappa H_x y,
\end{equation}

\noindent with $\kappa  ={2\pi d_e \mu _0}/{\Phi_0}$. Eq.(\ref{py})
leads to the well known Fraunhofer-like magnetic diffraction
pattern\cite{Jos1964}:

\begin{equation}
\label{Fraunhofer} I_c(H_x)= I_0 \left|\frac{\sin \pi H_x/H_{c}}{\pi H_x/H_{c} }\right|,
\end{equation}

\noindent where $I_0=4J_JWL$ is the zero field junction critical
current and $H_{c}=\Phi_0/2 \mu_0 d_e L$ is the so-called
\textit{(first) critical field}, i.e. the smallest field value for
which the Josephson current vanishes. Barone and Patern$\grave{\rm
o}$\cite{Barone} generalized Eq.(\ref{Fraunhofer}) to the case of an
arbitrary orientation of the external magnetic field in the junction
plane, ${\bf H_a} = {\rm H_x} {\bf \hat x}+ {\rm H_y} {\bf \hat y}$.
In such a case, still ${\bf H}={\bf H_a}$ and the resulting magnetic
diffraction pattern will be:

\begin{equation}
\label{Paterno} I_c(H_x,H_y)= I_0 \left| \frac{\sin \pi
H_x/H_{cx}}{\pi H_x/H_{cx}} \times \frac{\sin \pi H_y/H_{cy}}{\pi
H_y/H_{cy}} \right|,
\end{equation}

\noindent with $H_{cx}=\Phi_0/2 \mu_0 d_e L$ and $H_{cy}=\Phi_0/
2 \mu_0 d_e W$. Unfortunately, the last equation cannot be easily
generalized to the case of an arbitrary applied field orientation
${\bf H_a} = {\rm H_x} {\bf \hat x}+ {\rm H_y} {\bf \hat y}  + {\rm
H_z} {\bf \hat z}$, simply because, when $H_z \neq 0$, then ${\bf H}
\neq {\bf H_a}$. The effect of a transverse magnetic field has been
first considered in 1975 by Hebard and Fulton\cite{Hebard} in order
to provide a correct interpretation to some experimental data
published in the same year\cite{Rosenstein}. They observed that a
transverse applied field $\bf{H_a}= \rm{H_z} \bf{\hat{z}}$ induces
Meissner surface demagnetizing currents $\bf{j_s}$ feeding the
interior of the junction and so generating a magnetic field $\bf{H}$ in the
barrier plane such that $\bf{j_s} = \bf{\hat{z}} \times \bf{H}$. The problem
of finding the $\bf{j_s}$ distribution in a single superconducting
field subjected to a transverse magnetic field has been analytically
solved for different film geometries\cite{Landau,Brandt}. However, a
planar JTJ is made by two overlapping superconducting films
separated by a thin dielectric layer and the Meissner current
distributions on the interior surfaces (top surface of the bottom
film and bottom surface of the top film) require numerical approaches even for the more tractable electrode
configurations. Recently\cite{JAP08}, the magnetic field distribution
$\bf{H^\bot}$ in the barrier of small planar JTJs has been
numerically obtained in the case when an external magnetic field is
applied perpendicular to the barrier plane, ${\bf H_a}\equiv (0,0,H_z)$. The
simulations allowed for heuristic analytical approximations for the
Josephson static phase profile $\phi^\bot$ from which the dependence
of the maximum Josephson current $I_c(H_z)$ on the applied field
amplitude was calculated for the most common electrode geometrical
configurations (overlap, cross and annular junctions).
Unfortunately, the theoretical findings could not be tested against
experimental results due to the insufficiency of data available in
the literature.

One of the aims of this paper is to fill this vacancy. We have
measured the transverse magnetic diffraction patterns of several planar JTJs
with the most common geometrical configurations and compared the
results with their expected counterparts. More generally, we have
recorded the $I_c(H_a)$ when the applied field is oblique, that is, 
has non-zero in-plane and transverse components. To avoid
complications and without loss of generality, we have chosen the
in-plane component to be along one of the electrode axis, more
specifically, along the $x$-direction:

\begin{equation}
\label{Happl}  \bf{H_a}= \rm{H_x} \bf{\hat{x}}+ \rm{H_z} \bf{\hat{z}},
\end{equation}

\noindent so that, as shown in Fig.\ref{Geometry}, the applied
field forms an angle $\alpha$ with respect to the $x$-$y$ plane, that is:

\begin{equation}
{H_x}=H_a \cos\alpha \,\,\,\, \rm{and} \,\,\,\, {H_z}=H_a \sin\alpha,
\label{trigon}
\end{equation}

\noindent with $H_a= \sqrt{H_x^2+H_z^2}$. We will demonstrate that
the experimental oblique magnetic diffraction patterns can be nicely
reproduced by properly extending the theoretical framework of
Ref.\cite{JAP08}. This paper is constructed as follows. In
Sec.\ref{sec:Samples} we will present the samples used for the
measurements and describe the experimental setup. Sec.\ref{sec:Meas}
will report on the experimental results obtained for those samples whose
barrier has a rectangular shape (overlap-type junctions in
Sec.\ref{sec:OverJun} and cross-type junctions in
Sec.\ref{sec:CrossJun}). Sec.\ref{sec:AnnJun} will be devoted to the annular JTJs. Then, in
Sec.\ref{sec:Theory} we will discuss how to generalize the
theoretical analysis of the effect of a transverse magnetic field to the case
of an oblique field. Finally, the discussion and the interpretation
of the measurements will be given in Sec.\ref{sec:Discussion}, while the
conclusions will be presented in Sec.\ref{sec:Conc}.

\section{The samples\label{sec:Samples}}

High quality $Nb/Al-Al_{ox}/Nb$ JTJs were fabricated on $0.35\,mm$
thick silicon substrates using the trilayer technique in which the
junction is realized in the window opened in a $SiO_2$ insulator
layer - details of the fabrication process can be found in
Ref.\cite{VPK}. The so called passive or \textit{idle} region, i.e.
the distance of the barrier borders to the electrode borders, was on
the order of $1$-$2\,\mu m$ for all the junctions. The thickness of
the $SiO_2$ insulator layer was $400\,nm$. The demagnetization
currents strongly depend on the electrode thicknesses relative to
the London penetration depth. For our samples the nominal thicknesses of the base,
top and wiring $Nb$ layers were $200$, $100$ and $500\,nm$,
respectively. Considering that the London penetration depth for $Nb$
film is $\lambda_L \simeq 90\, nm$\cite{Broom}, we see that our samples
satisfy the thick film approximation. For all samples the high
quality has been inferred by a measure of the I-V characteristic at
$T=4.2\,K$. In fact, the subgap current $I_{sg}$ at $2\,mV $ was
small compared to the current rise $\Delta I_{g}$ in the
quasiparticle current at the gap voltage $V_{g}$, typically $\Delta
I_{g}>20I_{sg}$; the gap voltage was as large as $V_{g}=2.8\,mV$.
The geometrical and electrical (at $4.2\,K$) parameters of the seven
samples quoted in this paper are listed in Table I. For the
rectangular junctions $\#A$-$F$, beside their dimensions $2L$ and
$2W$ along the $x$ and $y$-directions, respectively, we also report
the junction aspect ratio $\beta=L/W$. (As shown in
Ref.\cite{JAP08}, this geometrical parameter turns out to be crucial
for the magnetic field line distribution in the barrier of a JTJ
subjected to a transverse magnetic field.) All the samples belonged to the
same fabrication batch (except sample $\# C$). Let us observe that
for the overlap-type junctions $\# A$ and $\# B$ the zero field
critical current $I_0$ was as large as the theoretical value $0.7\Delta
I_{g}$ predicted for strong-coupling $Nb$-$Nb$ JTJs, indicating the
absence of self-field effects. The critical current density has been
calculated as\cite{Cristiano} $J_c=0.7\,\Delta I_{g}/A$ in which
$\Delta I_{g}$ is the measured quasiparticle current step at the gap
voltage and $A$ is the junction nominal area ($A=2L\times2W$ for
rectangular junctions and $A=\pi (r_o^2-r_i^2)$ for the annular
junction). The Josephson critical current density was
$J_c=3.9kA/cm^2$ for all samples, except for sample $\# C$ having
$J_c=80A/cm^2$. The values of the barrier magnetic thickness $d_e =2
\lambda_L\ \simeq 180\,nm$ has been used to calculate the Josephson
penetration depth $\lambda _{J}=\sqrt{\phi_0 /2 \pi \mu _{0}d_{e}
J_{c}}$. (In the thin film limit, $\lambda _{J}$ can be better
determined  by using the expression for $d_e$ found by Weihnacht\cite{wei}.)
Accordingly, all samples had $\lambda_J \simeq 6\, \mu m$, except
sample $\# C$ which had $\lambda_J \simeq 42\, \mu m$. In other
words, as far as the electrical length concerns, all samples can be
classified as intermediate length junctions ($2L$,$2W\simeq
\lambda_J$), except sample $\# C$ that is a long ($2L>> \lambda_J$)
unidimensional ($2W<< \lambda_J$) overlap-type JTJ.

\begin{table}[bt]
    \centering
        \begin{tabular*}{0.85\textwidth}
{@{\extracolsep{\fill}}cccccccccc}
$JJ$&$geometry$&$2L\times2W$&$\beta$&$I_0$&$\Delta I_g$&$\Delta_{R}^{\|}$&${\Delta_{R}^{\bot}}$&$\eta_R$&$\alpha_M$\\
$\#$ &          & $\mu m^2$ & $L/W$ & $mA$   &  $mA$      & $\mu T$         &$\mu T$&$\Delta_{R}^{\bot}/\Delta_{R}^{\|}$&$$\\
\hline  
$A$ &$overlap$ &$10\times10$& $1$  & $3.9$  &  $5.6$     & $1290$           &$1100$&$0.85$&$140^\circ$\\
$B$ &$overlap$ &$5\times20$ &$0.25$& $3.9$  &  $5.6$     & $550$            &$190$&$0.35$&$160^\circ$\\
$C$ &$overlap$&$4\times500$&$0.008$& $1.0$  &  $2.3$     & $12$             &$0.88$&$0.073$&$176^\circ$\\
$D$ &$overlap$ &$20\times5$ & $4$  & $3.6$  &  $5.6$     & $2940$           &$5400$&$1.84$&$107^\circ$\\
$E$ &$cross$   &$10\times10$& $1$  & $3.8$  &  $5.8$     & $1080$           &$810$&$0.75$&$--$\\
$F$ &$cross$   &$20\times5$ & $4$  & $3.0$  &  $5.4$     & $420$            &$2240$&$5.3$&$--$\\
\hline
\hline
$JJ$&$geometry$&$r_i$  &$r_o$  &$I_0$&$\Delta I_g$&$\Delta_{A}^{\|}$&${\Delta_{A}^{\bot}}$&$\eta_A$&$\alpha_M$\\
$\#$ &          &$\mu m$&$\mu m$& $mA$   & $mA$       & $\mu T$      &$\mu T$&$\Delta_{A}^{\bot}/\Delta_{A}^{\|}$&$$\\
\hline  
$G$  &$annular$& $5$   & $8$   & $4.3$  &  $6.4$     & $490$       &$310$&$0.63$&$145^\circ$\\
        \end{tabular*}
    \caption{Relevant electrical (at T=4.2K) and geometrical parameters of the rectangular and annular $Nb/Al_{ox}/Nb$ Josephson tunnel junctions quoted in this paper. The Josephson critical current density was $J_c=3.9\,kA/cm^2$ (corresponding to $\lambda_J \simeq 6\,\mu m$) for all samples, except for sample $\#C$ having $J_c=80\,A/cm^2$ ($\lambda_J \simeq 42\,\mu m$). The experimental results obtained for these samples will be presented in Sec.\ref{sec:Meas} as follow: overlap-type junctions in Sec.\ref{sec:OverJun}, cross-type junctions in Sec.\ref{sec:CrossJun} and annular junctions in Sec.\ref{sec:AnnJun}.}
    \label{tab:Table1}
\end{table}

\begin{figure}[htb]
        \centering
                \includegraphics[width=9cm]{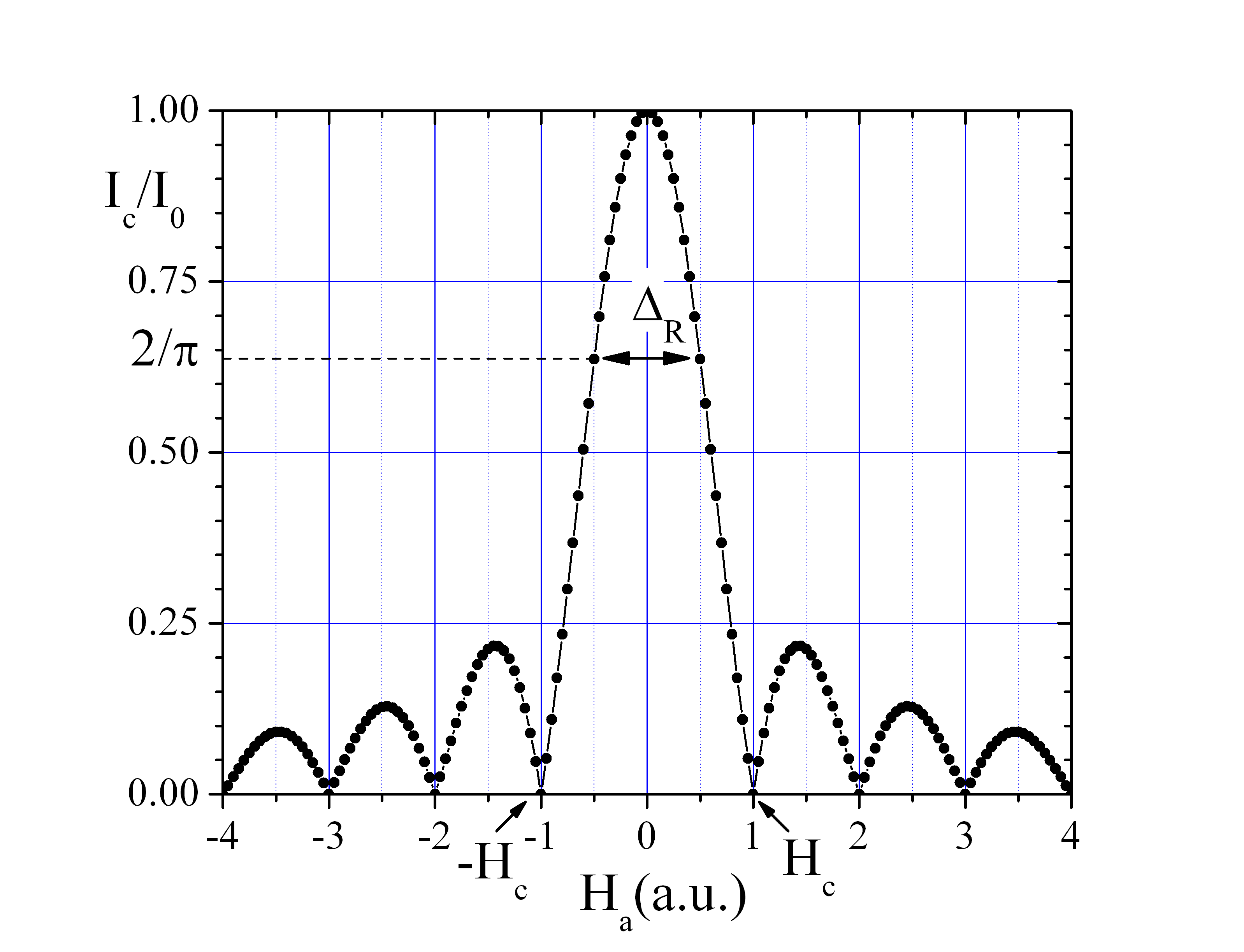}
\caption{(Color online) Definition of the parameter $\Delta_R$ as the
width of the magnetic field range in which $I_c(H_a)\geq (2/\pi)I_0$. By definition, for a Fraunhofer-like magnetic diffraction pattern, $\Delta_R$ coincides with the junction (first) critical field $H_c$.}
        \label{Delta}
\end{figure}

We now come to the definition of the parameters $\Delta_{R}^{\|}$,
${\Delta_{R}^{\bot}}$ and their ratio $\eta_R$ whose experimental values are reported in Table
I for the rectangular junctions ($\Delta_{A}^{\|}$,
${\Delta_{A}^{\bot}}$ and $\eta_A$ for the annular junction). As already
mentioned in the Introduction, it's well known that the magnetic
diffraction pattern of an electrically small rectangular JTJ in
the presence of an in-plane field perpendicular to one of the
barrier edge follows the Fraunhofer pattern in Eq.(\ref{Fraunhofer})
characterized by a periodic amplitude modulation. As depicted in
Fig.\ref{Delta}, the value $H_c$ of the applied field where
first the critical current vanishes is called the (first) critical
field. ($H_c$ is a measure of the response (for small field) of the
junction critical current to the applied field: the smaller is
$H_c$, the larger is the junction response.)
For those samples whose $I_c(H_a)$ is still amplitude modulated, but
follows a different pattern (for example, annular, circular and
rhombic junctions), the critical field $H_c$ can still be defined as
that value of external field $H_a$ where first the critical current
nulls, $I_c(H_c)=0$. Further, for those samples whose $I_c(H_a)$
shows modulation lobes, but never vanishes (as examples, small
junctions with nonuniform tunneling current and long JTJs), the
critical field can still be obtained extrapolating to zero the first
modulation lobe. However, in those cases in which the critical
current $I_c$ is a monotonically decreasing function of the applied
field $H_a$, the concept of critical field $H_c$ looses its meaning
and a new feature has to be introduced to characterize the behavior
of $I_c(H_a)$ for small fields. A theoretical example is offered by a
Gaussian shaped junction subjected to an in-plane magnetic  field
that is characterized by a Gaussian magnetic diffraction
pattern\cite{Peterson}. A practical example is given by a square
cross junction in a transverse field, whose $I_c(H_a)$ decreases
with $H_a$ with no measurable modulation\cite{Miller85}. The
experimental magnetic diffraction patterns that will be reported in
the next section span all kinds of behaviors from Fraunhofer-type to
$1/H_a^\nu$-like (with $\nu>0$). Therefore, as the new and universal
figure of merit to characterize the response of the critical current
to the externally applied field amplitude, we have chosen the width
of the magnetic field range $\Delta_R$ in which $I_c(H_a)\geq
(2/\pi)I_0 \simeq 0.64I_0$ (see Fig.\ref{Delta}). Considering
that, when Eq.(\ref{Fraunhofer}) holds, $I_c(H_c/2)=(2/\pi)I_0$, the
value of the prefactor stems from the requirement that the new merit
figure $\Delta_R$ numerically equals the critical field $H_c$
whenever the measured magnetic pattern follows a Fraunhofer
dependence, i.e., $\Delta_R=H_c$. In all other cases, generally
speaking, $\Delta_R\neq H_c$. In our notation, $\Delta_{R}^{\|}$ and
${\Delta_{R}^{\bot}}$ are the merit figures of, respectively, an
in-plane ($\alpha=0$) and transverse ($\alpha=90^\circ$) magnetic
diffraction pattern. With a similar reasoning, we define the
parameter $\Delta_A$ for annular junctions as the width of the
magnetic field range $\Delta_R$ in which $I_c(H_a)\geq \mu I_0$,
with $\mu \simeq 0.67$. The slightly different prefactor stems from
the fact that the in-plane diffraction pattern of a small annular
junction (with no trapped fluxon) follows a Bessel-type
dependence\cite{br96}: $I_c(H_a)= I_0 \left|J_0(\zeta_1
H_a/H_c)\right|$ in which $J_0$ is the zero-order Bessel function
and $\zeta_1\approx 2.405$ is its first zero. Now $H_c=\Phi_0/ \mu_0
d_e C$, where $C$ is the ring mean circumference. 

\noindent The measurement of $\Delta$ requires an external field smaller than the one required for $H_c$; henceforth, this new parameter also
turns out to be a very useful quantity whenever the junction
critical field cannot be experimentally determined since it exceeds
the irreversible field, i.e., when the Abrikosov vortices first enter into
the superconducting films and become pinned in the
junction\cite{Finnemore}. For our samples the transverse
irreversible field was about $5\, mT$ ($50\, Gauss$).

\noindent The ratios $\eta_R= \Delta_{R}^{\bot}/\Delta_{R}^{\|}$ and
$\eta_A= \Delta_{A}^{\bot}/\Delta_{A}^{\|}$ provide a direct
comparison between the $I_c$ response to a transverse field relative
to the in-plane field; specifically, $\eta<1$ means that the
junction critical current modulates faster when the applied field is
transverse. In a recent paper\cite{JAP07} we already provided an
experimental proof that a transverse magnetic field can be much more
capable than an in-plane one to modulate the critical current $I_c$
of a planar JTJ with proper barrier and electrodes geometry
requirements. This property was first obtained and exploited in the
context of a detailed investigation of the phase symmetry breaking
during fast normal-to-superconducting phase transitions of long
annular JTJs\cite{PRL06}.

Our setup consisted of a cryoprobe inserted vertically in
a commercial $LHe$ dewar. The cryoprobe was magnetically shielded by
means of two concentric magnetic shields: the inner one made of $Pb$
and the outer one of cryoperm. Inside the vacuum tight can of the
cryoprobe, a non-magnetic insert holds a chip mount with spring
contacts to a $Si$ chip with planar JTJs. With reference to the coordinate system in Fig.\ref{Geometry}, the chip was positioned in the center of a long
superconducting cylindrical solenoid whose axis was along the
$x$-direction (within less than $1^\circ$ of accuracy) to provide an
in-plane magnetic field. In order to provide a transverse magnetic
field, a superconducting cylindrical coil was placed $5\, mm$ far
from the chip with its axis oriented along the $z$-direction (within
less than $3^\circ$ of accuracy) . Two independent low-noise dc current sources
were used to feed the solenoid and the coil in order to expose our
samples at magnetic fields having arbitrary magnitude and
orientation (in the $x$-$z$-plane). The field-to-current ratio was
$3.9\, \mu T/mA$ for the solenoid and $4.4\, \mu T/mA$ for the coil.
These values have been numerically obtained from Comsol Multiphysics
magnetostatic simulations in order to take into account the strong
correction to the free-space solution due to the presence of the
close fitting superconducting shield\cite{morten}.

\begin{figure}[ht]
        \centering
                \includegraphics[width=8cm]{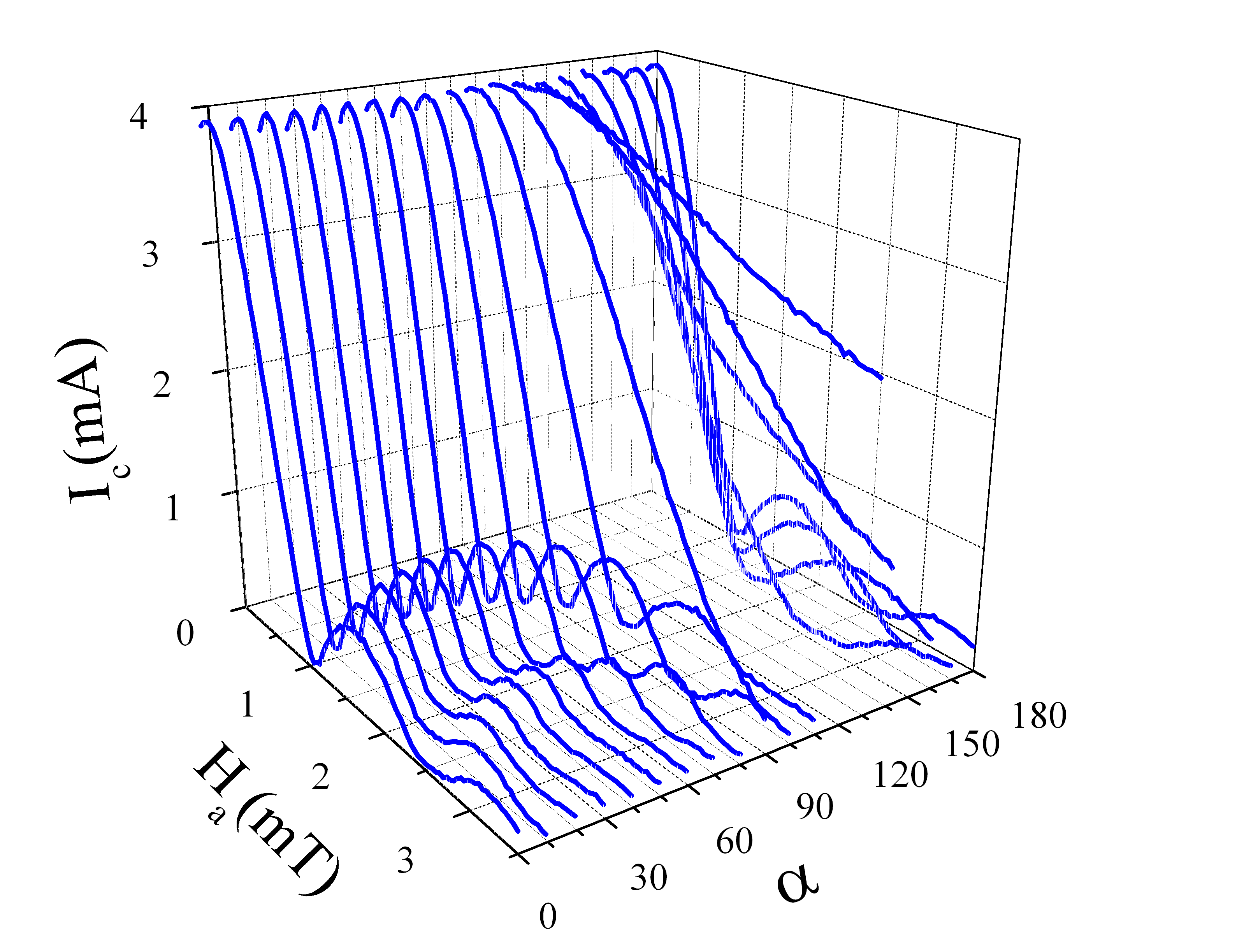}
        \caption{(Color online) Tridimensional plot showing for the square overlap junction ($\beta=1$) the recorded magnetic diffraction patterns $I_c(H_a)$ for different values of the field orientation $\alpha$ (with $\Delta \alpha = 10^{\circ}$). }
        \label{JJC}
\end{figure}

\section{The measurements\label{sec:Meas}}

In this Section we present the experimental oblique magnetic
diffraction patterns relative to planar JTJs having the seven
different electrode configurations listed in Table I. Subsection A, B and C will concern samples having, respectively, overlap, inline and annular geometry. The theoretical interpretation of our data-sets will be given in the next Section.
The angle $\alpha$ that the external oblique field forms with
barrier plane could be experimentally spanned in the interval $[-\pi, \pi]$; however, according to Eq.(\ref{trigon}), an angle rotation
of $\pm \pi$ is equivalent to an inversion of the field direction,
i.e., of the field amplitude $H_a \rightarrow -H_a$. For this reason
we will only present data for $\alpha$ in the $[0, \pi]$ interval
with the amplitude $H_a$ assuming both negative and positive values.
Further, by denoting with $I_{c}^+$ and $I_{c}^-$ the positive and negative
critical currents, respectively, we always had $I_{c}^-(H_a)=I_{c}^+(-H_a)$, as expected, due to the absence of any measurable stray fields in our setup. For this reason, we will only present data for $I_{c}^+$, which we will simply call $I_{c}$. We stress that, in recording the $I_c$ vs. $H_a$ curves, we took special care
that the applied field never exceeded the reversible field, so it
was not expected that the applied field penetrated the films. Furthermore,
through measurements of the sample's I-V characteristic, it was
verified that $H_a$ was so small as not to affect the energy gap.
Finally, the raw experimental data were postprocessed to take into
account the difference in the solenoid and coil field-to-current
factors.

\begin{figure}[htb]
\centering
\subfigure[]{\includegraphics[width=7cm]{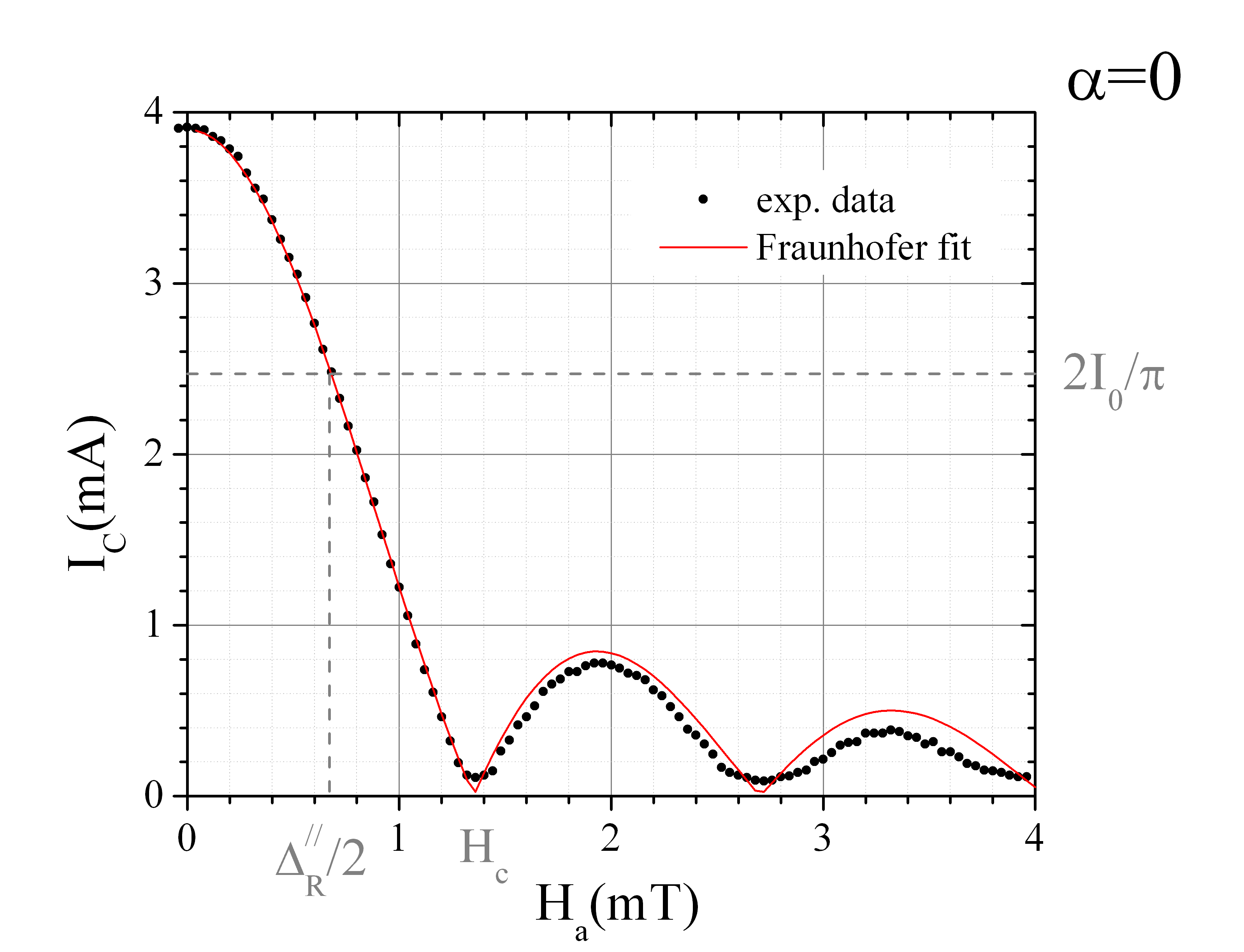}}
\subfigure[]{\includegraphics[width=7cm]{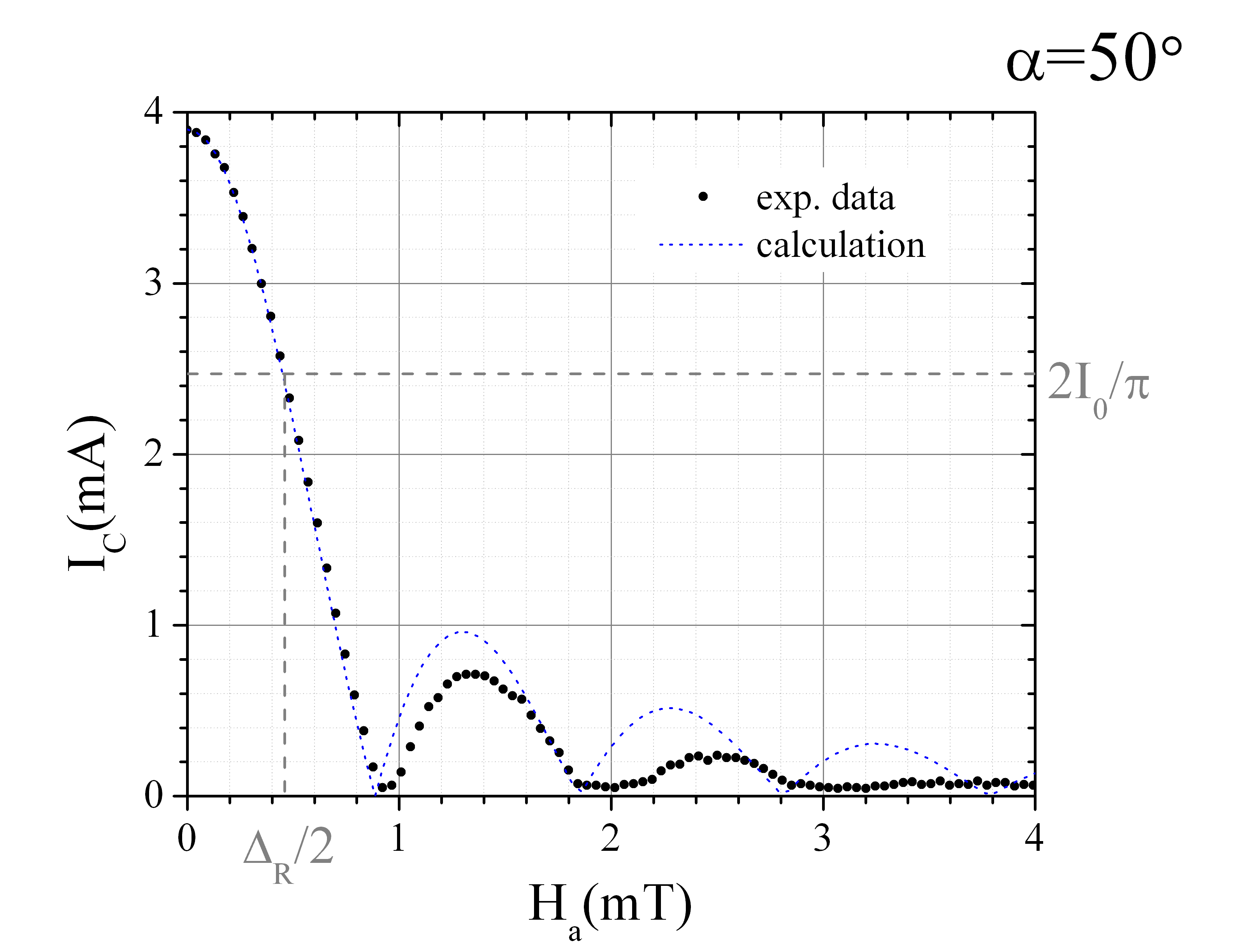}}
\subfigure[]{\includegraphics[width=7cm]{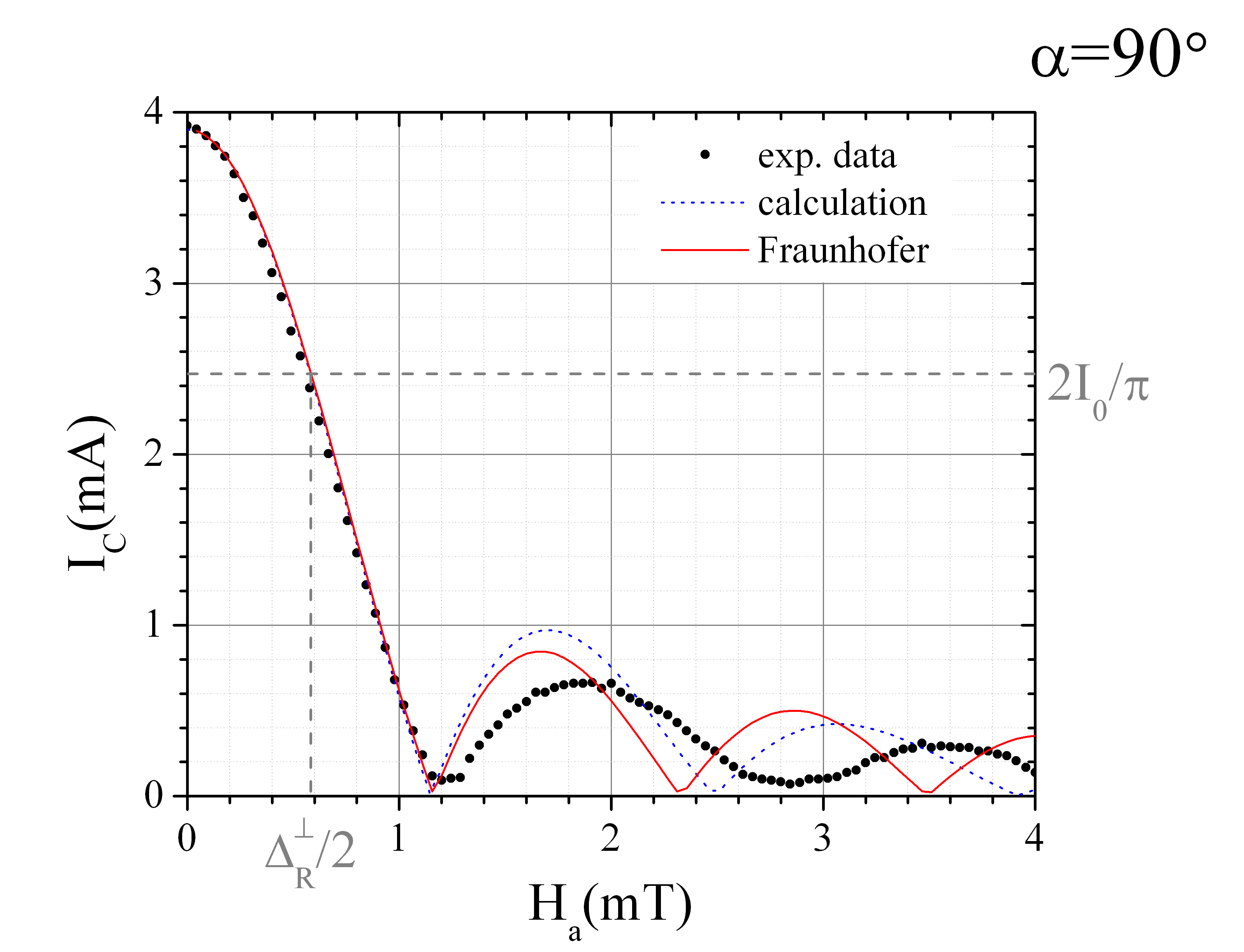}}
\subfigure[]{\includegraphics[width=7cm]{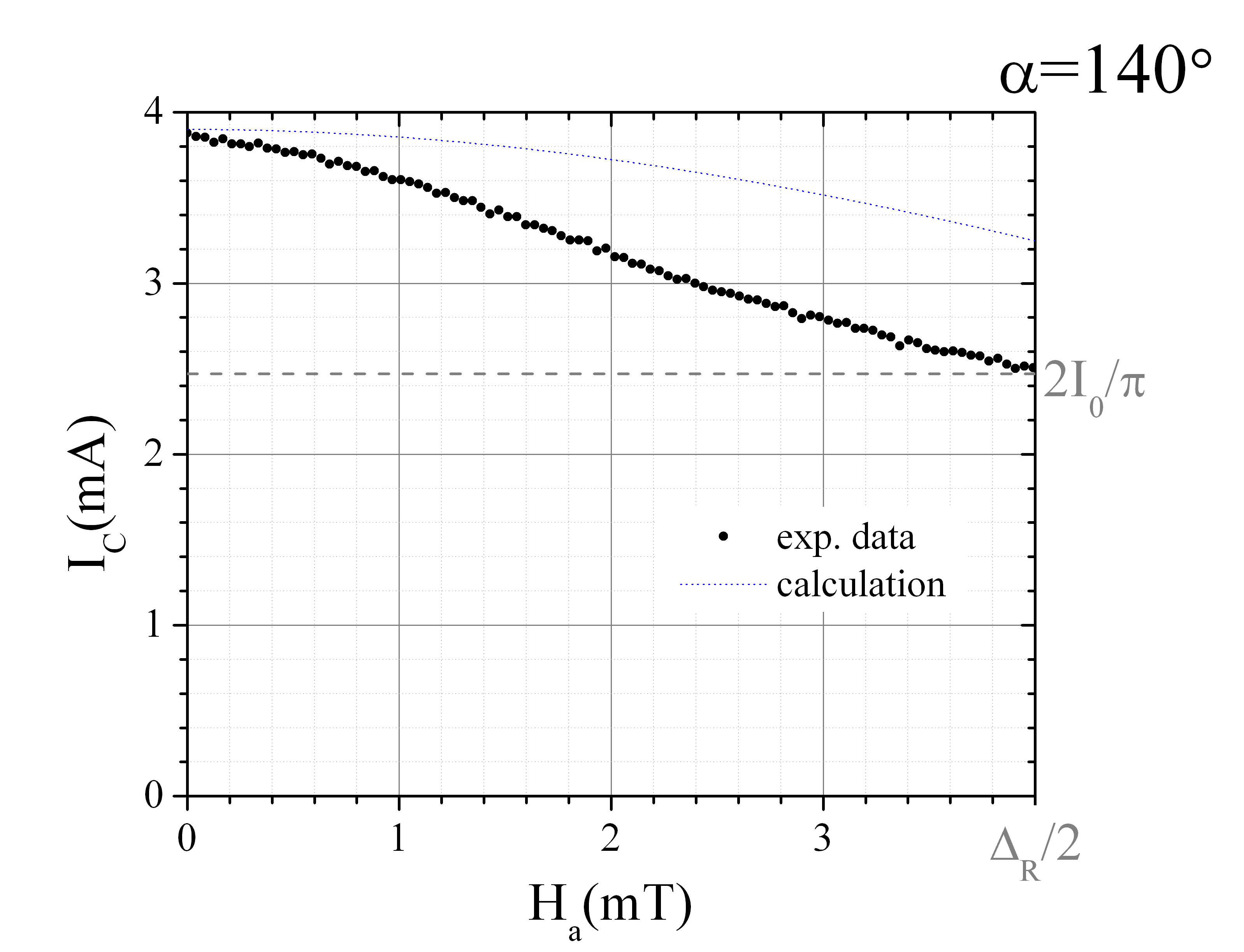}}
			\caption{(Color online) Magnetic diffraction patterns of the square
overlap junction ($\beta=1$) for different $\alpha$ values: a)
$\alpha=0$ (in-plane field), b) $\alpha=50^{\circ}$, c)
$\alpha=90^{\circ}$ (transverse field), and d) $\alpha=140^{\circ}$. The
experimental data are presented by closed circles. The solid lines,
when present, are the best Fraunhofer fit, while the dotted lines are
the results of the calculations described in Sec.\ref{sec:Theory}. For
$\alpha=0$, by construction, the calculations reproduce the Fraunhofer shape.}
\label{JJCa-b}
\end{figure}

\subsection{Overlap-type junctions\label{sec:OverJun}}

We begin with an intermediate length, square overlap-type JTJ,
namely sample $\#A$ in Table I ($2L=2W \simeq 1.6 \lambda_J$). Fig.\ref{JJC} is a tridimensional plot of the magnetic diffraction
patterns recorded for different $\alpha$ values (with $\Delta \alpha
= 10^{\circ}$). Since, for this sample, $I_{c}(-H_a)=I_{c}(H_a)$, we
only show the data for $H_a\geq 0$.  It is evident that $I_c(H_a)$
smoothly, but drastically, changes with the field orientation
$\alpha$. To be clearer, in Figs.\ref{JJCa-b} we report the
magnetic patterns for four selected $\alpha$ values. For $\alpha =0$
the applied field is in the barrier plane ($z=0$) and, as seen in
Fig.\ref{JJCa-b}a, $I_c(H_a)$ closely follows a Fraunhofer-like
behavior, as expected. The small discrepancy between the
experimental data (closed circles) and the Fraunhofer fit (solid
line) can be ascribed to the fact that junction dimensions are
slightly larger than the Josephson penetration depth. Increasing
$\alpha$, in the beginning the junction critical field $H_c$ (or
equivalently the width of the pattern main lobe $\Delta_R$ defined
earlier) first slowly decreases until it reaches an absolute
minimum when $\alpha \simeq 50^{\circ}$ (see Fig.\ref{JJCa-b}b),
and later on quickly increases until it reaches an absolute maximum
when $\alpha \simeq 140^{\circ}$ (see Fig.\ref{JJCa-b}d.) In
Fig.\ref{JJCa-b}c we also report the transverse ($\alpha =
90^{\circ}$) magnetic pattern to evidence how much it differs from a
Fraunhofer dependence. It is worth stressing that,
whenever $\alpha\neq 2m \pi$ (with integer $m$), the magnetic
diffraction pattern looses the modulation periodicity
$H_{cn}=nH_{c1}$ featuring the Fraunhofer behavior; more
specifically, the distance between two adjacent minima increases as
we move to larger fields, i.e., $H_{cn}>nH_{c1}$. Each plot in Figs.\ref{JJCa-b} explicitly reports the corresponding position of
the measured $\Delta_R$.

\begin{figure}[ht]
\centering
\subfigure[]{\includegraphics[width=6cm]{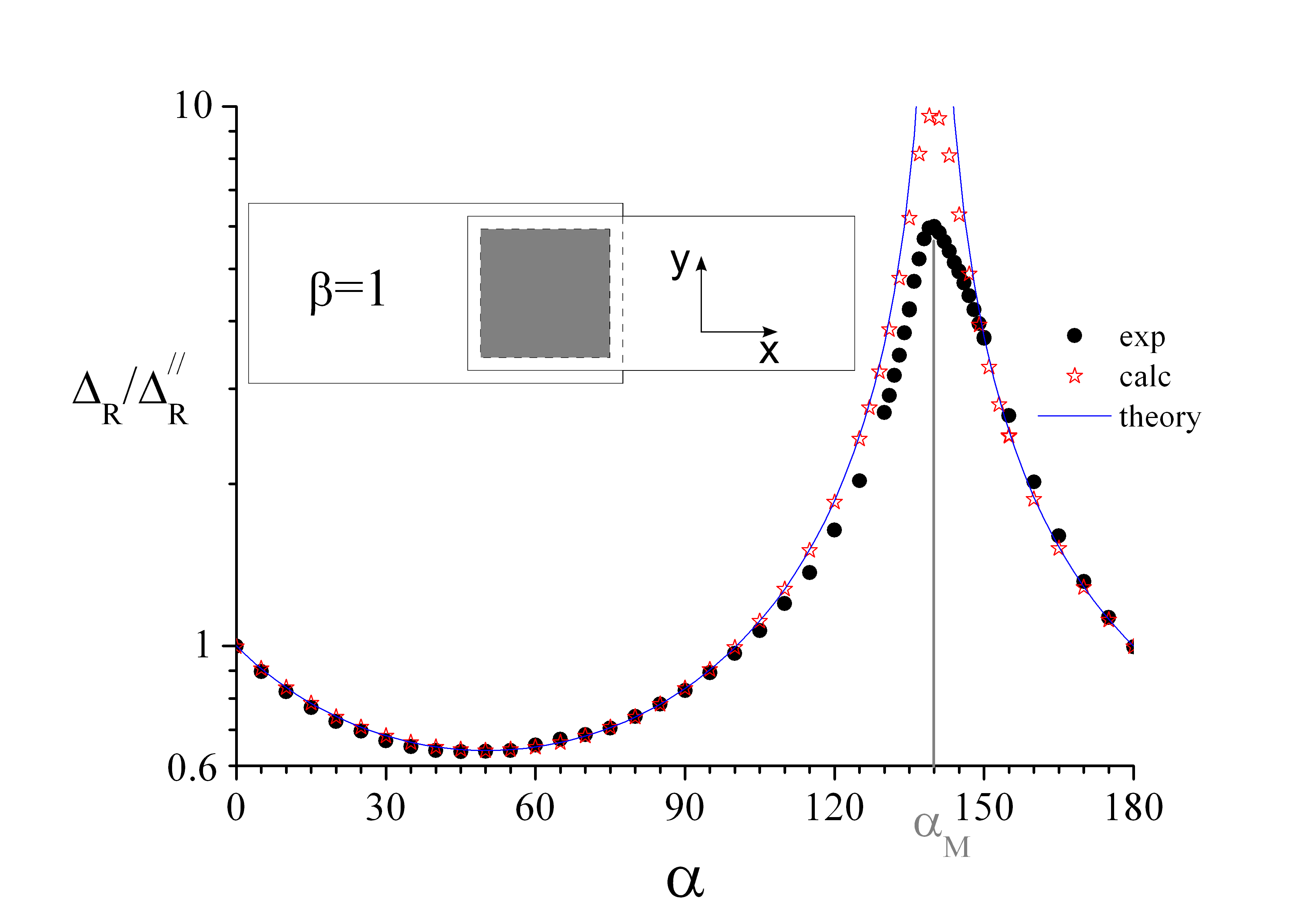}}
\subfigure[]{\includegraphics[width=6cm]{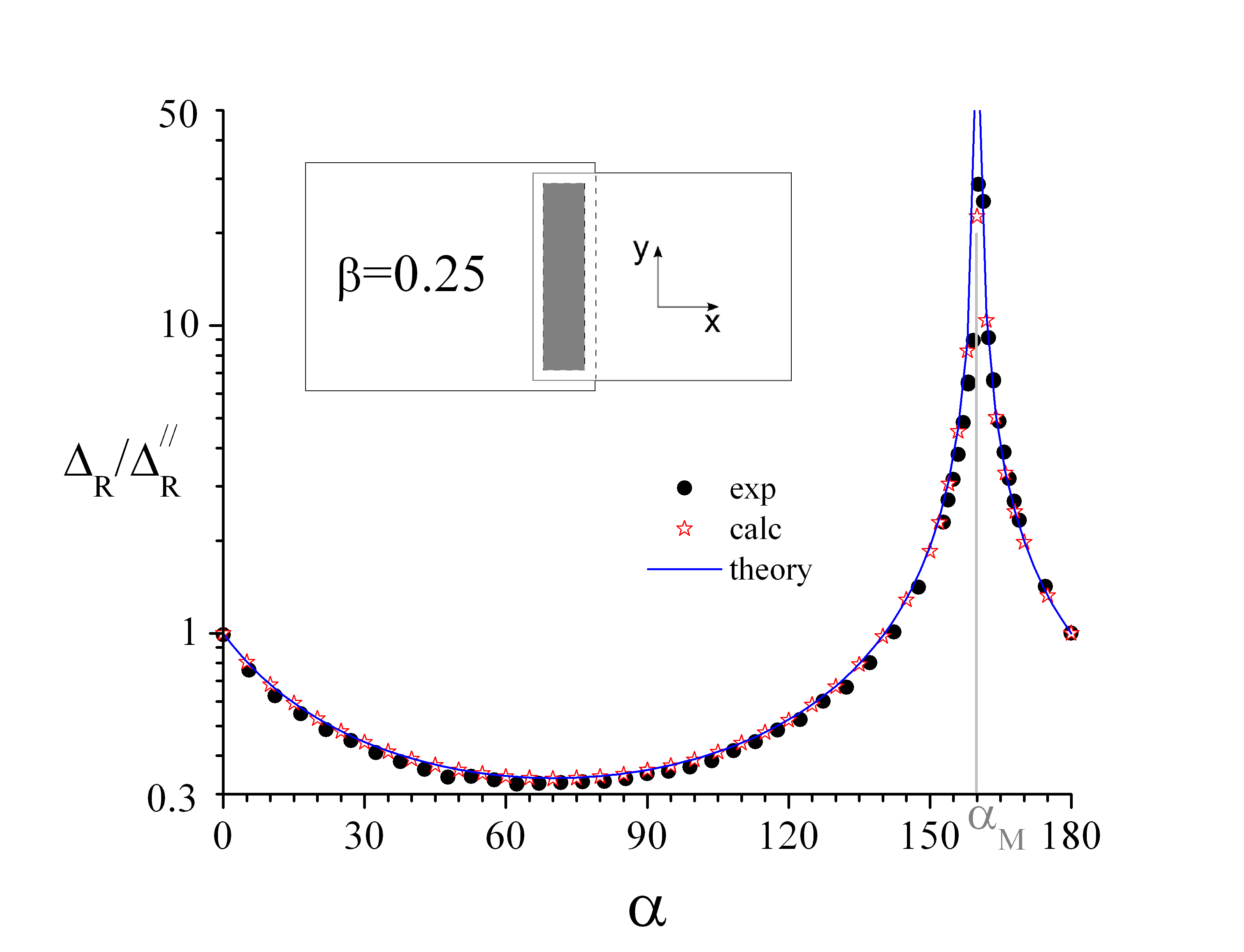}}
\subfigure[]{\includegraphics[width=6cm]{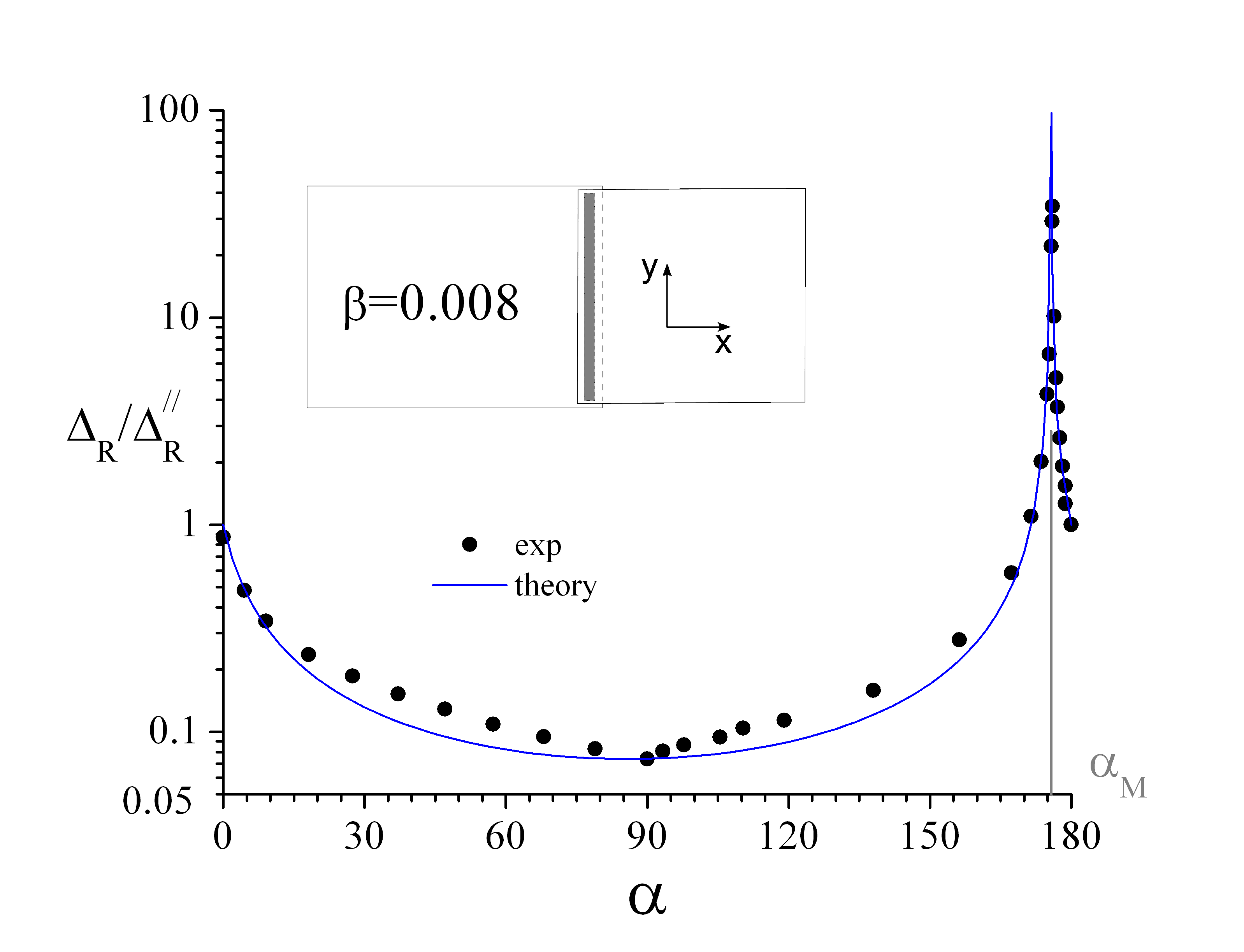}}
\subfigure[]{\includegraphics[width=6cm]{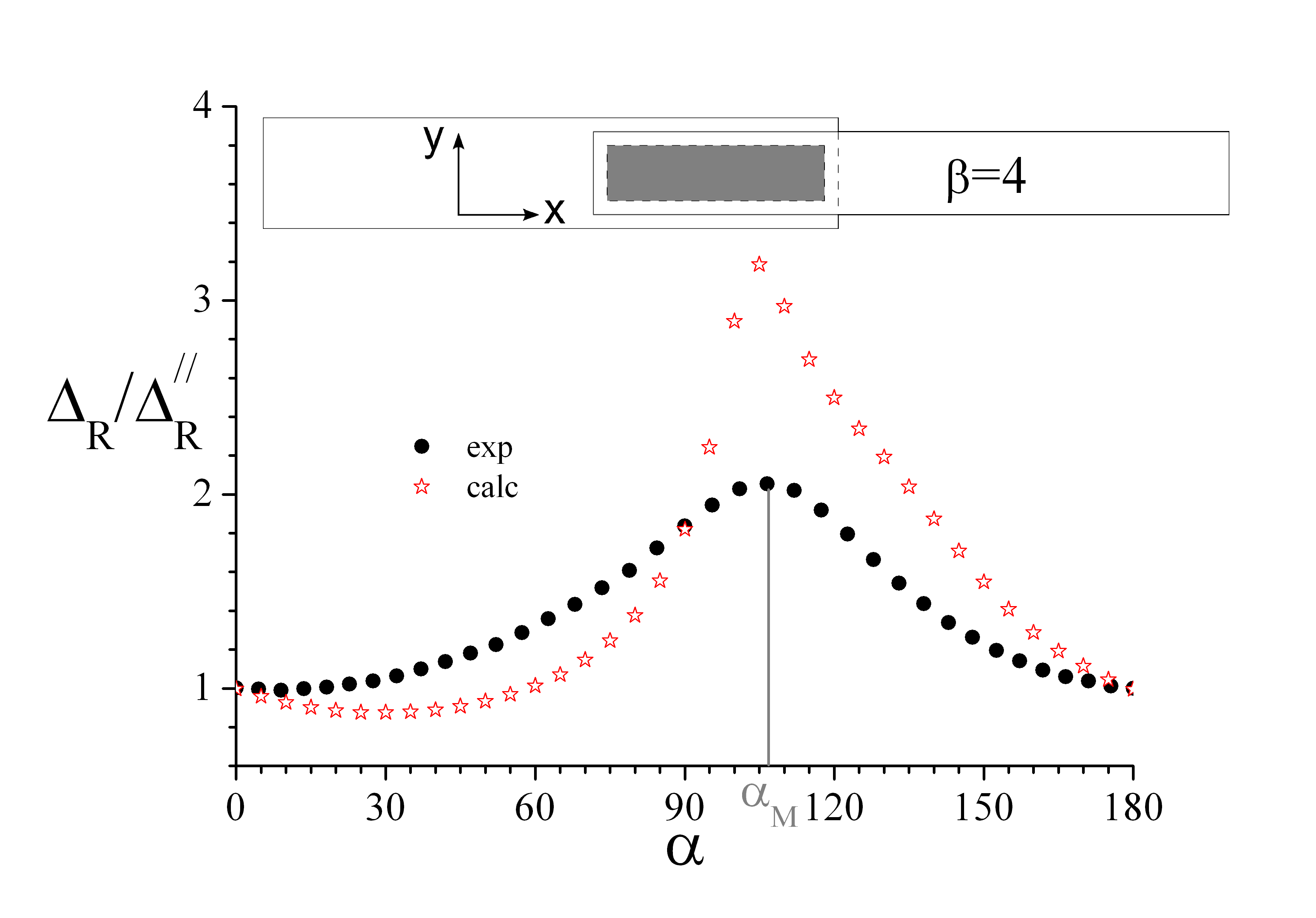}}
			\caption{(Color online) Magnetic field range $\Delta_R$ vs. $\alpha$ (in degrees) for overlap junctions with different aspect ratios $\beta$: a) $\beta=1$, b) $\beta=0.25$, c) $\beta=0.08$, and d) $\beta=4$. The experimental data are presented by closed circles. The open stars result from the calculations described in
Sec.\ref{sec:Theory}, while the solid lines, when present, are the
result of a simple theory developed in Sec.\ref{sec:Discussion}. The
insets sketch for each sample its electrode configuration and its
orientation with respect to the chosen Cartesian coordinates.}
\label{OverDeltaAlfa}
\end{figure}

\noindent The $\alpha$ dependence of $\Delta_R$ (normalized to
$\Delta_R^{\|}$) for the sample $\#A$ is summarized in Fig.\ref{OverDeltaAlfa}a (solid circles). $\Delta_R(\alpha)$ is
reported in  Figs.\ref{OverDeltaAlfa}b-d for the samples $\#B$,
$\#C$, and $\#D$, overlap-type JTJs having aspect ratios,
respectively, $\beta=0.25$, $0.08$, and $4$. The insets in
Figs.\ref{OverDeltaAlfa}a-d sketch for each sample its electrode
configuration and its orientation with respect to the Cartesian
coordinates chosen in the Introduction (see Fig.\ref{Geometry}). We used a vertical log-scale for those samples having $\beta \leq 1$. Each plot in Figs.\ref{OverDeltaAlfa} is characterized by an absolute maximum achieved when
$\alpha=\alpha_M$. The $\alpha_M$ values, quoted in the last column
of Table I, were found to monotonically depend on the $\eta_R$
ratios which, in turn, scale with the $\beta$ ratios. In
Sec.\ref{sec:Discussion} we will discuss a simple theoretical
approach aimed to find the $\alpha$ dependence of $\Delta_R$ and the
relationship between $\alpha_M$ and $\eta_R$, as well.

We like to point out that the only measurements similar to those
reported in Fig.\ref{OverDeltaAlfa} can be found in a
pioneering paper dated 1975 by Rosenstein and Chen\cite{Rosenstein}.
They measured the first and second junction critical fields in an
oblique magnetic field for an overlap-type planar JTJ having
$\beta\simeq 0.5$ (and formed by two $300\, nm$ thick $Pb$
electrodes of unequal widths). They found that both $H_{c1}$ and
$H_{c2}$ reach their maximum values when the field orientation is
about $8^{\circ}$ off the in-plane direction ($\alpha_M \simeq
172^{\circ}$ in our notation). This value is consistent with our
findings.

\subsection{Cross-type junctions\label{sec:CrossJun}}

\begin{figure}[b]
        \centering
                \includegraphics[width=6cm]{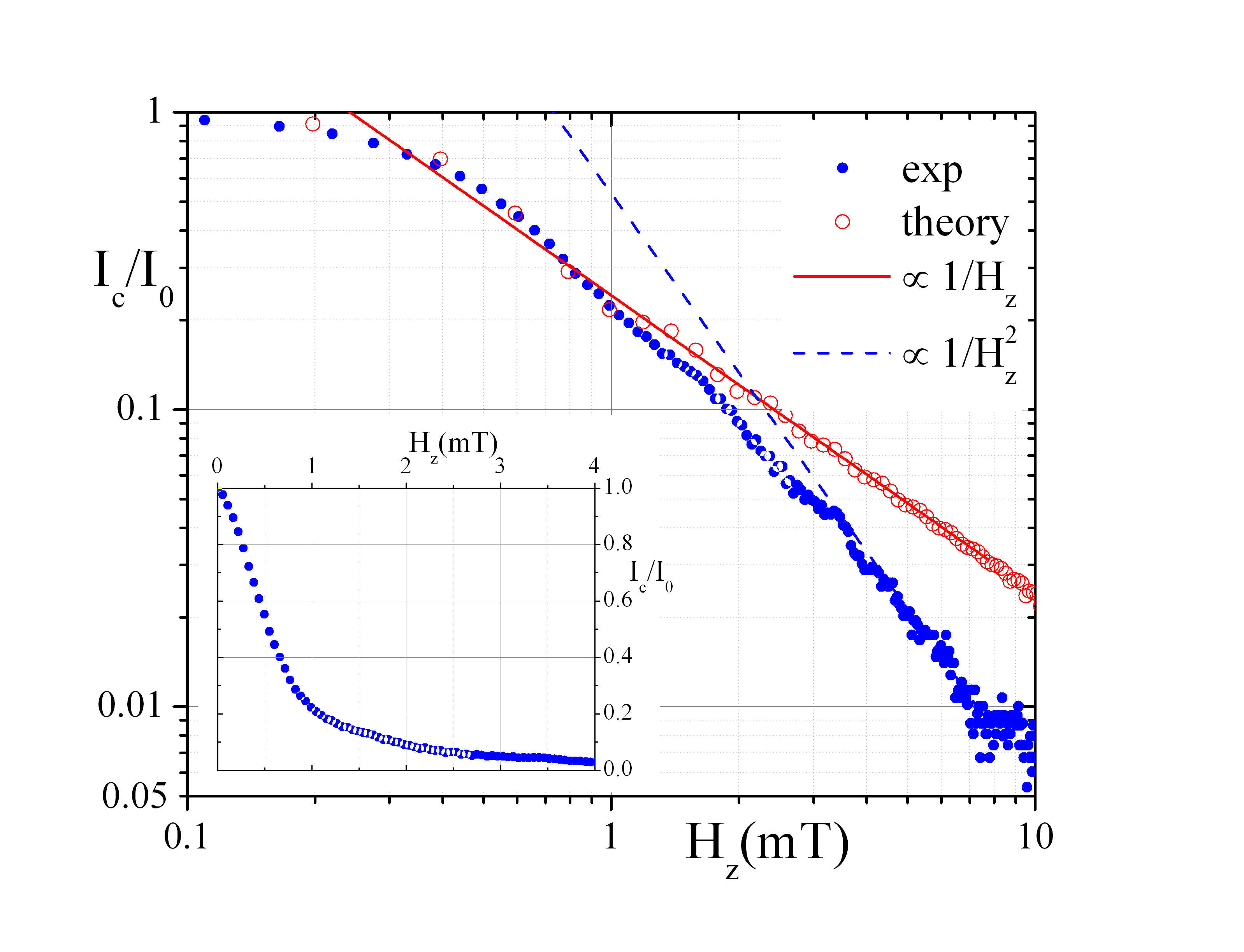}
        \caption{(Color online) Log-log graph of the transverse magnetic pattern $I_c(H_z)$ of the cross square junction. The closed squares are the experimental data, while the open circles are the result of computations based on (\ref{phicross}). The dashed and solid lines are the large-field best fit of the experimental and computed data, respectively, $\propto H_z^{-2}$ and  $\propto H_{z}^{-1}$. The inset shows the experimental data on linear scales.}
        \label{Crosstransverse}
\end{figure}

Figure \ref{Crosstransverse} shows the transverse magnetic
pattern of the square cross junction (sample $\#E$ in Table I) on a
log-log plot. The inset displays the same data on linear scales. We
observe that the critical current monotonically decreases as the
external field is increased. The experimental data indicate that for
large fields, $I_c \propto H_{z}^{-2}$, in contrast with the simple
inverse proportionality suggested by Miller \textit{et
al.}\cite{Miller85}.

\begin{figure}[t]
        \centering
 \subfigure[]{\includegraphics[width=7cm]{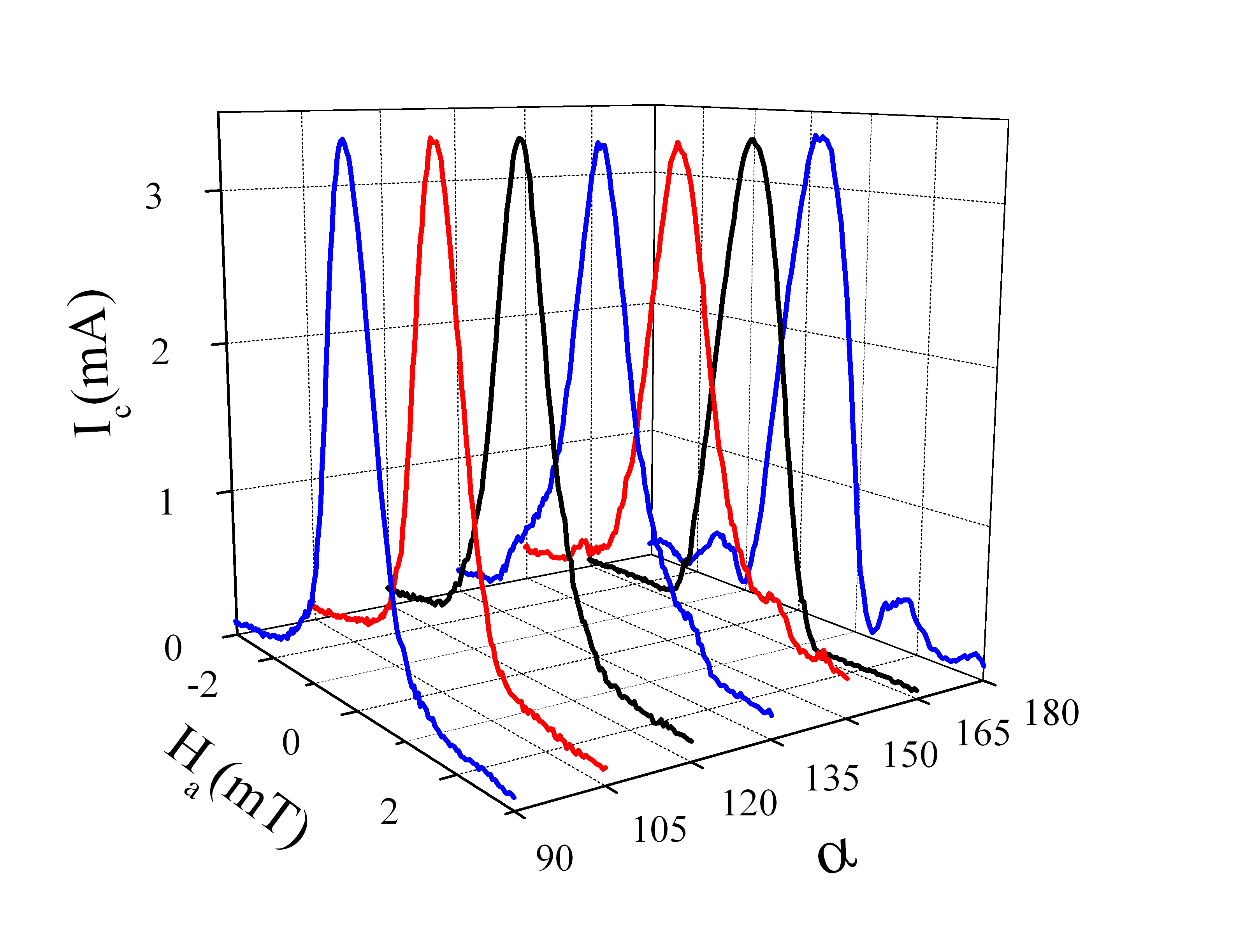}}
 \subfigure[]{\includegraphics[width=7cm]{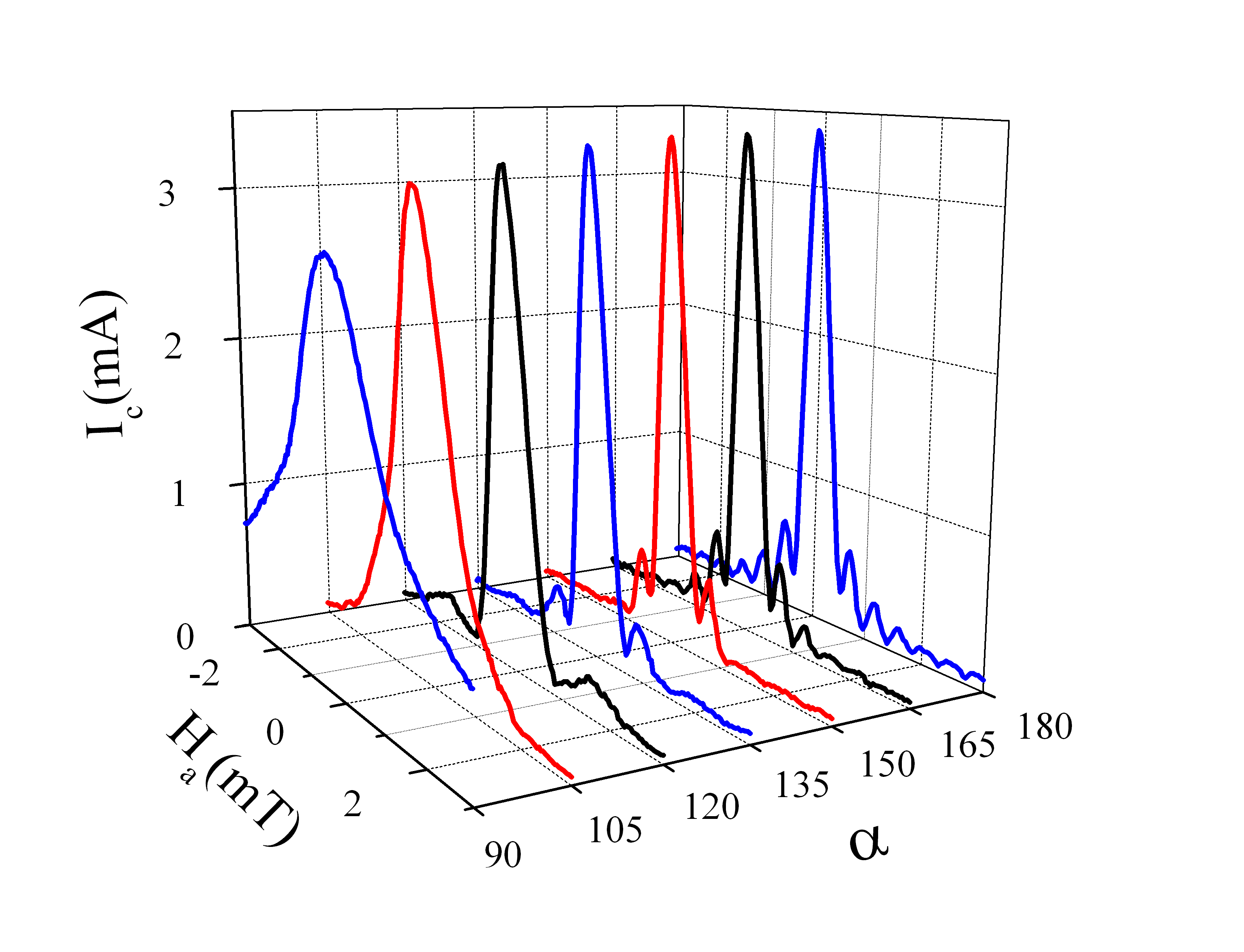}}
        \caption{(Color online) Tridimensional plots for the recorded oblique magnetic diffraction patterns $I_c(H_a, \alpha)$ of two cross-type junctions: a) square junction with $\beta=1$ and b) asymmetric junction with $\beta=0.25$. The angular separation is $\Delta \alpha = 15^{\circ}$. (For these samples $I_c(H_a, \pi/2+\alpha)= I_c(-H_a, \pi/2-\alpha)$.)}
        \label{JJAB}
\end{figure}

\begin{figure}[b]
\centering
\subfigure[]{\includegraphics[width=7cm]{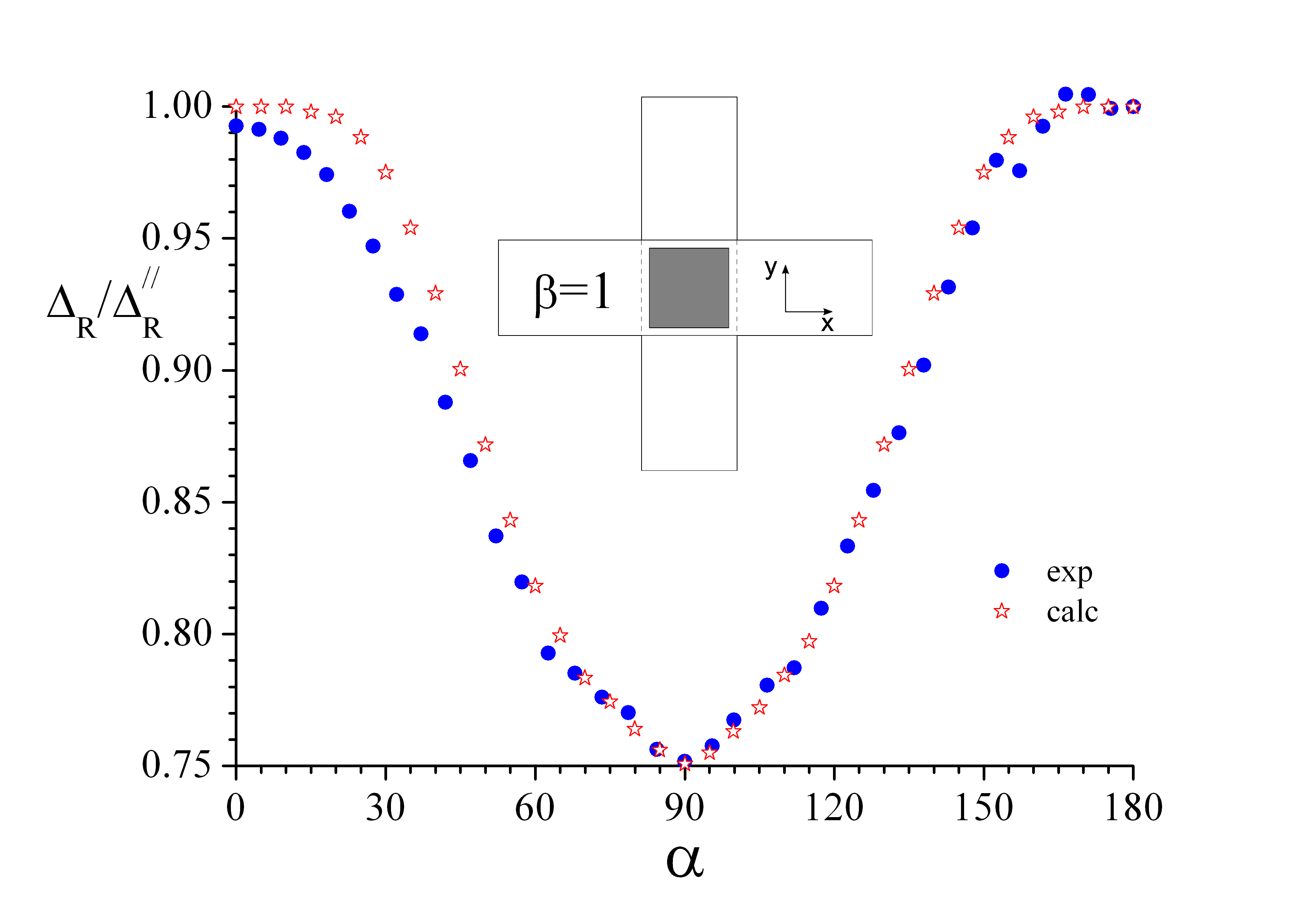}}
\subfigure[]{\includegraphics[width=7cm]{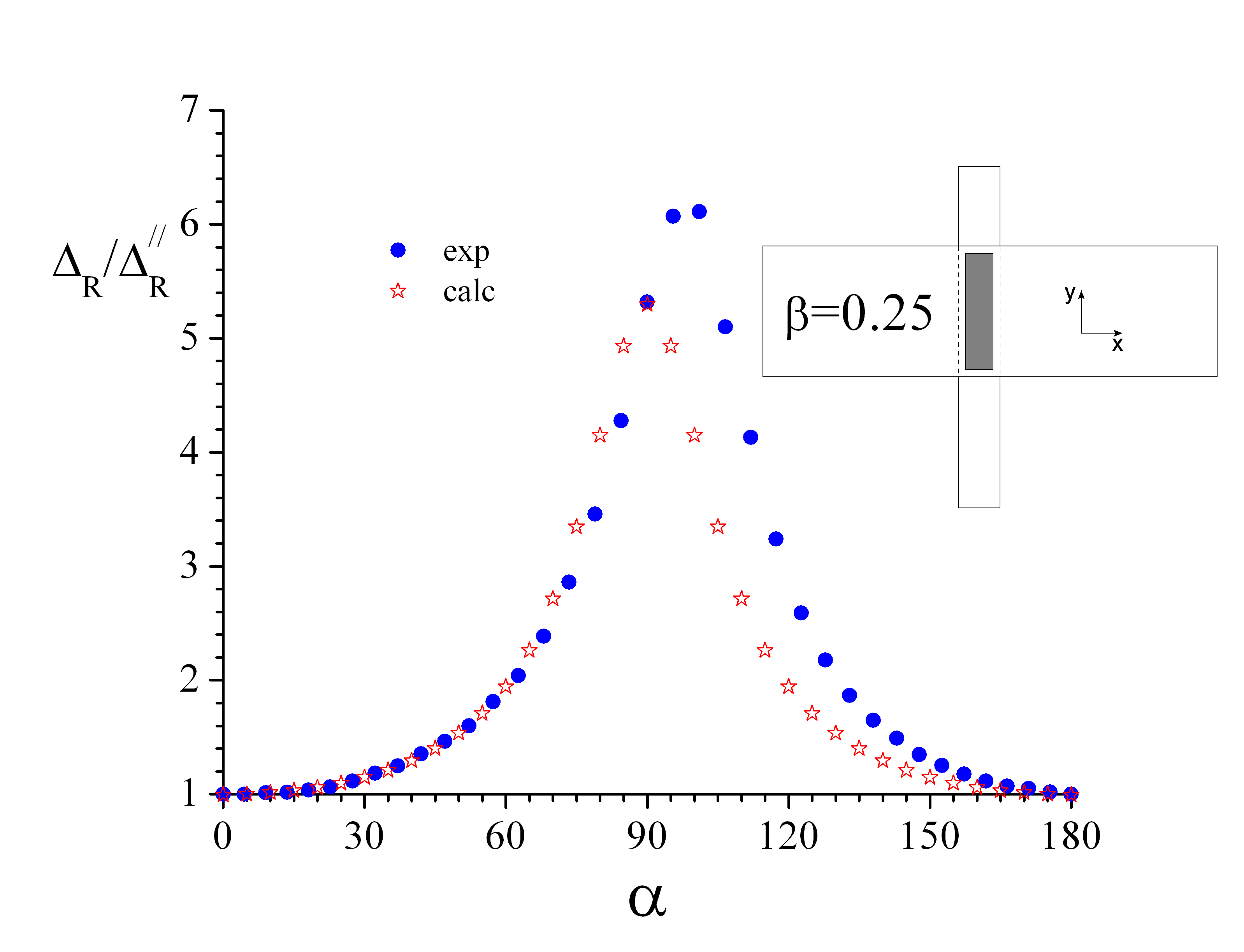}}
			\caption{(Color online) Magnetic field range $\Delta_R$ vs. $\alpha$
(in degrees) for the two cross-type junctions of Figs.\ref{JJAB}: a) $\beta=1$ and b) $\beta=0.25$. The experimental data are presented by closed circles, while the open stars
are the result of the calculations described in Sec.\ref{sec:Theory}. The
insets sketch for each sample its electrode configuration and its
orientation with respect to the chosen Cartesian coordinates.}
\label{CrossJunctions}
\end{figure}

\noindent In Figs.\ref{JJAB}a) and b) we report the oblique
magnetic diffraction patterns of the two cross-type junctions quoted in Table
I, respectively, $JJ\#E$ and $JJ\#F$. For these samples the in-plane
patterns are skewed due to the self-field effects. The skewness is
more pronounced for the asymmetric sample, having $2L\approx 3
\lambda_J$ and $I_0=I_c(H_a=0)=3.6\, mA$ (of course $I_0$ does not
depend on $\alpha$). As we move from an in-plane field (say
$\alpha=0$) to a transverse one (say $\alpha=90^\circ$), the
skewness gradually disappears and,  keeping increasing  $\alpha$ toward
$180^\circ$, the skewness changes its polarity; in other words,
$I_c(H_a, \pi/2+\alpha)= I_c(-H_a, \pi/2-\alpha)$. For this reason,
we only present the $I_c(H_a, \alpha)$ plots for $\alpha$ in the
$[\pi/2, \pi]$ range (with $\Delta \alpha = 15^{\circ}$). The two
samples show a quite different $\alpha$ dependence of the normalized
pattern width $\Delta_R/\Delta_R^{\|}$, as shown in Figs.
\ref{CrossJunctions}. While the former one is characterized by a
weak dependence of $\Delta_R (\alpha)$ with a minimum when the
applied field is transverse, the latter one shows a substantial
variation with a maximum when the applied field is close to be
transverse, more specifically, when $\alpha \simeq 105^\circ$.

\subsection{Annular junctions\label{sec:AnnJun}}

\begin{figure}[t]
        \centering
                \includegraphics[width=8cm]{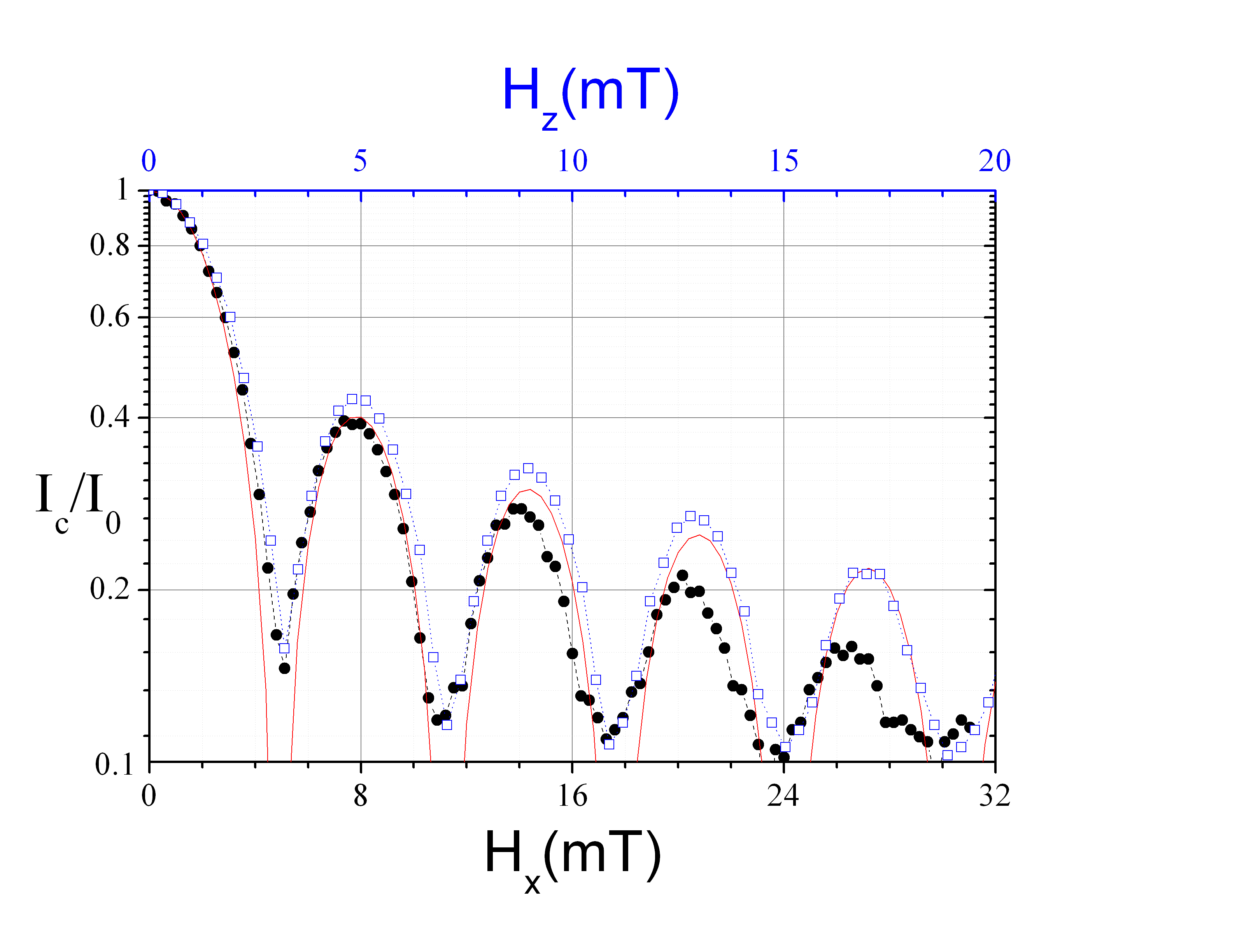}
        \caption{(Color online) Comparison of the magnetic patterns recorded for the annular junction quoted in Table I ($\bar{r} \approx \lambda_J$) in a transverse (open squares referred to top horizontal scale) and in-plane (closed circles referred to the bottom horizontal scale) applied magnetic field. The logarithmic vertical scale helps the data comparison in the lower current range. The solid line corresponds to a Bessel-like fit according to Eq. (\ref{IcAnn}).}
        \label{annularComparison}
\end{figure}

In two recent papers\cite{JAP07,JAP08}, among other things, we have
reported on the transverse magnetic diffraction patterns of
ring-shaped $Nb$-based annular JTJs with radii ten times (or more)
larger than the Josephson penetration depth. In this section we
present the results relative to a sample having the mean radius
$\bar{r} \approx \lambda_J$, namely $JJ\#G$ in Table I. Fig.\ref{annularComparison} compares in a combined plot the
transverse and the in-plane  recorded magnetic diffraction patterns:
respectively, the open squares referred to top horizontal scale
($H_z$) and the closed circles referred to the bottom horizontal
scale ($H_x$). The vertical logarithmic scale was needed to enhance
the plot differences that would be otherwise barely observable using
a vertical linear scale (a part of the quite different horizontal
scales). On a first order of approximation both patterns closely
follow the zero-order Bessel function behavior; the solid line in
Fig.\ref{annularComparison} is the best data fit using
Eq.(\ref{IcAnn}) with the first critical field as a unique fitting
parameter. Fig.\ref{annularComparison} indicates that for this
particular sample a transverse field modulates the junction critical
current about $1.5$ times faster than an in-plane field. In
Ref.\cite{JAP07} we have shown that this gain increases with the
ring diameter and can be even larger than $100$.

\begin{figure}[b]
        \centering
                \includegraphics[width=8cm]{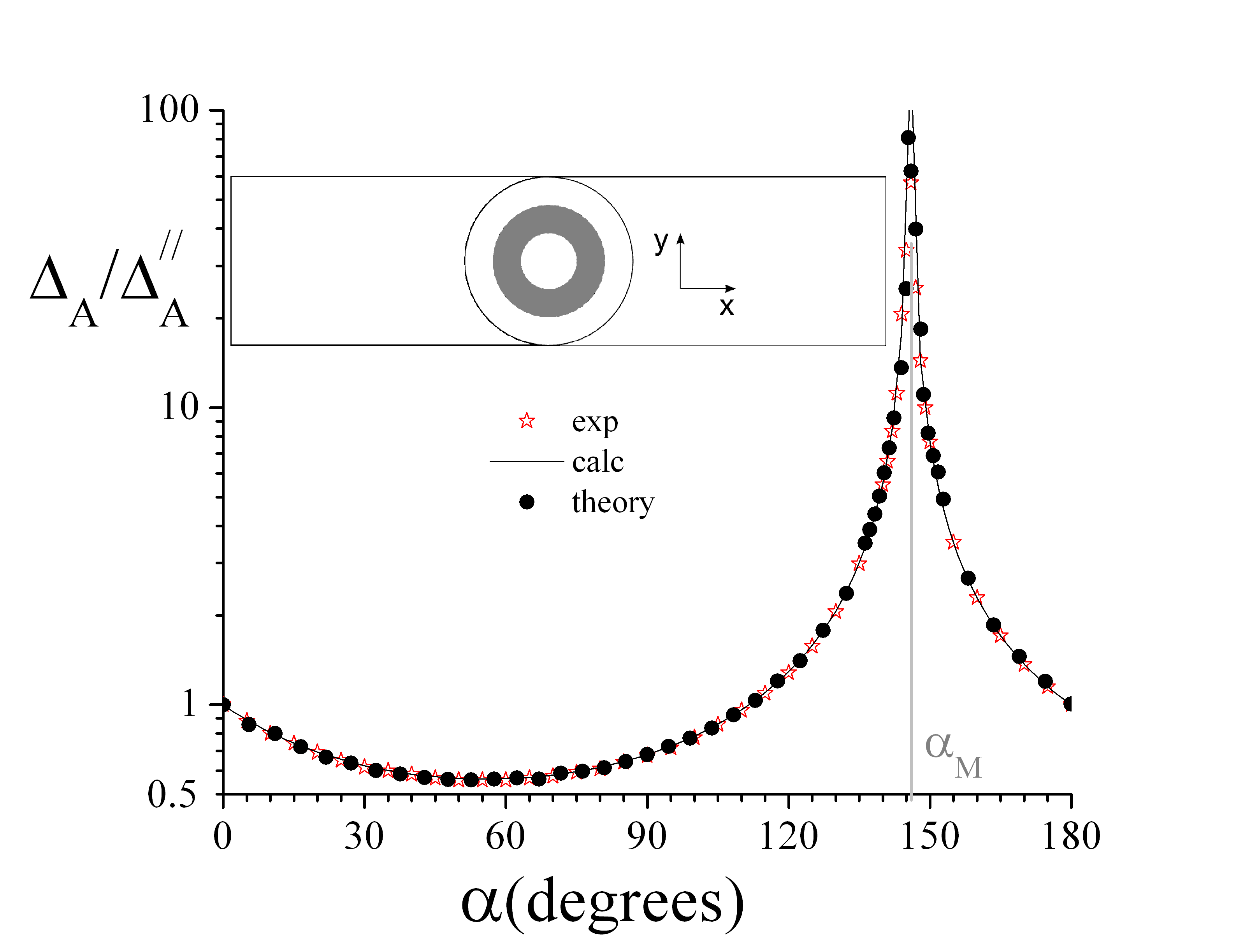}
        \caption{(Color online) Magnetic field range $\Delta_A$ vs. $\alpha$ (in degrees) for the annular junction quoted in Table I ($\bar{r} \approx \lambda_J$). The experimental data are presented by closed circles, while the open stars are the result of calculations described in Sec.\ref{sec:Theory}. The solid line arises from the approximate analytical expression Eq.(\ref{Deltalfa}). The inset sketches the \textit{Lyngby-type} annular junction and its orientation with respect to the chosen Cartesian coordinates.}
        \label{annular}
\end{figure}

\noindent As shown in Fig.\ref{annular}, the annular range
width $\Delta_A$ drastically depends on the external field
orientation $\alpha$. In fact, although its values for $\alpha=0$
and $90^\circ$ belong to the same order of magnitude,
$\Delta_{A}^{\bot} = 0.63 {\Delta_{A}^{\|}}$, we see that
$\Delta_A(\alpha)$ is peaked at $\alpha_M \simeq 145^\circ$, with
$\Delta_A(\alpha_M) \simeq  100 \Delta_{A}^{\|} \simeq  160
\Delta_{A}^{\bot}$; in other words, when $\alpha \simeq \alpha_M$,
the sample is practically insensitive to the external magnetic
field, and an external field amplitude as large as the irreversible
field is required to reduce the critical current $I_c$ to $67\%$ of
its zero field value $I_0$.

\section{Theory\label{sec:Theory}}

In order to provide a theoretical interpretation of the experimental
data presented in the previous Section, let us introduce the spatial
normalized units $\bar{x}=x/L$ and $\bar{y}=y/W$ with the junction
center coinciding with the axis origin. Our task is to find out the
Josephson phase distribution $\phi(\bar{x},\bar{y})$ over the
barrier area ($-1\leq \bar{x} \leq 1$ and $-1 \leq \bar{y} \leq 1$)
of a small planar JTJ in a weak oblique applied magnetic field ${\bf
H_a}$. As a preliminary step, before resorting to Eq.(\ref{gra}), we
need to determine the magnetic field distribution over the barrier
area ${\bf H}(\bar{x},\bar{y})$. As the Maxwell equations are linear
in the magnetic field, one can resort to the principle of
superposition to calculate the field ${\bf H}$. Thus, the effect of
the oblique field ${\bf H_a}$ can be conveniently split into the sum
of the effects of two orthogonal components, that is, the in-plane
component $\sqrt{H_x^2 + H_y^2}$ and the transverse one $H_z$. In
other words:

\begin{equation}
\label{Hsum} {\bf H}(\bar{x},\bar{y})={\bf H^\|}(\bar{x},\bar{y})+{\bf H^\bot}(\bar{x},\bar{y}),
\end{equation}

\noindent in which ${\bf H^\|}$ and ${\bf H^\bot}$ are the barrier
field distributions induced by in-plane and transverse external
fields, respectively. As mentioned in the Introduction, it has been
traditionally assumed that when the external field lays in the
barrier plane ($H_z=0$ and ${\bf H^\bot}=0$), it uniformly threads
the oxide layer, so that ${\bf H^\|}={\bf H}={\bf H_a} \equiv (H_x,
H_y)$. Today we know that this is only true to the first approximation
for \textit{naked} JTJs, since field focusing effects should to be considered
in planar JTJ structures especially in the case of window
junctions\cite{JAP95,Ustinov} that are surrounded by a passive thick
oxide layer, the so-called \textit{idle region}\cite{Caputo}. Our samples were
designed to have the smallest possible idle region, so that field
focusing effects could be neglected; consequently our theory has
been developed under the simplifying assumption that the samples are
\textit{naked}. Further, as shown in Fig.\ref{Geometry}, we will only
consider magnetic field directions confined to the plane specified
by the angle $\alpha$ between the applied field and the junction
plane ($H_y=0$). According to  Eq. (\ref{Hsum}), due to the
linearity of Eq.(\ref{gra}), also the phase distribution can be
written as the sum of two terms $\phi_\|(\bar{x},\bar{y})$ and
$\phi_\bot(\bar{x},\bar{y})$:

\begin{equation}
\label{phisum}  \phi(\bar{x},\bar{y})= \phi^\|(\bar{x},\bar{y}) + \phi^\bot(\bar{x},\bar{y}),
\end{equation}

\noindent provided that: ${\bf \nabla} \phi^\| = \kappa {\bf H^\|}\times {\bf \hat z}$ and
${\bf \nabla} \phi^\bot = \kappa {\bf H^\bot}\times {\bf \hat z}$.

\noindent Once $\phi(\bar{x},\bar{y})$ is known, it will be possible to
calculate the junction critical current $I_c$ as\cite{Miller85}:

\begin{equation}
\label{Ic} I_c= I_0 \sqrt{ \langle \sin \phi \rangle^2 + \langle \cos \phi \rangle^2},
\end{equation}

\noindent in which the brackets $\langle\rangle$ denote spatial
averages over the junction area - $4\langle f(\bar{x},\bar{y})
\rangle = \int_{-1}^{1} d\hat{x} \int_{-1}^{1} d\hat{y}
f(\bar{x},\bar{y})$.

Being ${\bf H^\|}=H_x {\bf \hat{x}}$ (for naked small JTJs),
${\phi^\|}$ is given by Eq.(\ref{py}) that, with our
normalization, becomes:

\begin{equation}
\label{phipar}  \phi^\|(\bar{x},\bar{y}) = h^\| \bar{y} \cos\alpha,
\end{equation}

\noindent in which $h^\|= 2\pi \mu_0 H_a W d_e  / \Phi_0 $, since
$H_a \cos\alpha = H_x$. (If one wants to consider the effect of the
idle region, still $h^\| \propto H_a$, but the proportionality
constant needs proper correction.) ${\bf H^\bot}$ and their
corresponding ${\phi^\bot}$ have been found in
Ref.\cite{JAP08} for naked small JTJs having the most common
electrode configurations. Generally speaking, it was found
${\phi^\bot} \propto H_z = H_a \sin\alpha$, so that we are allowed to
write:

\begin{equation}
\label{phiperp}  \phi^\bot(\bar{x},\bar{y}) = h^\bot  \phi_\bot(\bar{x},\bar{y}) \sin\alpha,
\end{equation}

\noindent with $h^\bot \propto H_a$ and $\phi_\bot(\bar{x},\bar{y})$
containing the spatial part of $\phi^\bot(\bar{x},\bar{y})$. It is important
to stress that the proportionality constant between $h^\bot$ and
$H_a$ is not known a priori, being a non-trivial and still unknown
function of several geometrical junction features such as the widths
and the thicknesses of the two electrodes, their separation and
their configuration\cite{JAP08}. Therefore $h^\bot$ remains
the only free parameter when comparing the
experimental data to their theoretical counterparts. Eq.(\ref{phisum}) can be now rewritten in terms of Eqs.(\ref{phipar}) and (\ref{phiperp}) as:

$$ \phi(\bar{x},\bar{y})= h^\| \bar{y} \cos\alpha + h^\bot  \phi_\bot(\bar{x},\bar{y}) \sin\alpha. $$

\noindent Being both $h^\|$ and  $h^\bot$ proportional to the
intensity of the applied field $H_a$, their ratio $\eta = h^\bot /
h^\|$ depends uniquely on geometrical parameters (including the
junction magnetic thickness $d_e$). Therefore, the last equation can
be conveniently cast in its final form:

\begin{equation}
\label{phisum1}  \phi_{h,\alpha}(\bar{x},\bar{y})= h \left[ \bar{y}
\cos\alpha + \eta  \phi_\bot(\bar{x},\bar{y}) \sin\alpha\right],
\end{equation}

\noindent in which, $h^\|$ has simply been renamed $h$ and the
indices $h$ and $\alpha$ have been added to explicitly indicate that
the Josephson phase distribution depends on the field strength and
orientation. In the remaining part of this Section, we will
extensively make use of Eq.(\ref{phisum1}), inserting, for each
junction geometrical configuration, the proper $\phi_\bot$
expression from Ref.\cite{JAP08}.

\subsection{Overlap-type junctions\label{sec:OverJun_T}}

From the numerical analysis of the magnetic scalar potential induced
in the barrier plane of a planar JTJ by a transverse field we were
able to derive approximate and simple expressions for the Josephson
phase distribution in the barrier area that satisfy the Laplace
equation (${\partial^2 \phi}/{\partial x^2} + {\partial^2
\phi}/{\partial y^2}=0$). These heuristic expressions were found to
be markedly dependent on the junction aspect ratio $\beta =L/W$. For
an overlap-type junction in a unitary transverse magnetic field we
found:

\begin{equation}
\label{phiover} \phi_{\bot}(\bar{x}, \bar{y}) \approx
\sin\bar{y}\frac{\cosh\beta \bar{x}}{\sinh\beta},
\end{equation}

\noindent Inserting Eq.(\ref{phiover}) into (\ref{phisum1}), we
observe that $\phi_{h,\alpha}$ is an odd function of $\bar{y}$,
henceforth $\langle \sin \phi_{h,\alpha} \rangle=0$; furthermore,
considering that $\phi_{h,\alpha}$ is an even function of $\bar{x}$,
the calculation Eq.(\ref{Ic}) of the magnetic diffraction pattern
reduces to:

\begin{equation}
\label{overpatt} I_c(h,\alpha)=I_0\int_{0}^{1} d\bar{x} \int_{0}^{1}
d\bar{y} \, \cos \phi_{h,\alpha} \, .
\end{equation}

\noindent The above integral has been numerically evaluated as a
function of the reduced field $h$ and for several values of $\alpha$, setting
$I_0=3.9\, mA$, $\beta=1$ and $\eta=0.85$ in order to make a
comparison with the experimental data for the square overlap
junction of Figs.\ref{JJCa-b} - the calculated $I_c$ vs. $h$ are
reported as dotted lines and a proper horizontal scaling was chosen to
match the in-plane field range $\Delta_{R}^{\|}=1.29mT$.

\noindent By construction, for $\alpha=0$, Eq.(\ref{overpatt})
returns the Fraunhofer pattern. For $\alpha \neq 0$ the calculated
$I_c(h)$, at a quantitative level, is only consistent with the
experimental data, the discrepancies being more evident for large
field values. However our calculations can grasp most of the pattern
small field features: in particular, they can nicely reproduce the
dependence of the pattern width $\Delta_R$ on $\alpha$, as far as
$\beta \leq 1$ (see the open stars in Figs.
\ref{OverDeltaAlfa}a-d). In fact, for $\beta=4$, to use
expression (\ref{phiover}) only allows us to reproduce the correct
value of $\alpha_M$.

\subsection{Cross-type junctions\label{sec:CrossJun_T}}

\noindent For a cross-type small naked JTJ with aspect ratio $\beta$,
it was heuristically found\cite{JAP08}:

\begin{equation}
\label{phicross} \phi_{\bot}(\bar{x}, \bar{y}) =
\sin\bar{y}\frac{\sinh\beta \bar{x}}{\cosh\beta} +
\sin\bar{x}\frac{\sinh \bar{y}/\beta}{\cosh1/\beta}.
\end{equation}

\noindent This approximate expression, beside satisfying the Laplace
equation, has the proper symmetry properties required by the
problem, $\phi^{\beta}_{\bot}(\bar{x},
\bar{y})=\phi^{1/\beta}_{\bot}(\bar{y},\bar{x})$. We will consider
first the relevant case of a square cross junction, i.e., $\beta=1$.
When this is the case, by retaining the first terms in the Taylor
expansion of the trigonometric and hyperbolic functions, the above
expression (\ref{phicross}) reduces to:

\begin{equation}
\label{Miller} \phi_{\bot}(\bar{x}, \bar{y}) = \bar{x} \bar{y} ,
\end{equation}


\noindent proposed by Miller \textit{et al.}\cite{Miller85}  in 1985.
The transverse magnetic pattern of a square cross junction can be
computed through Eq.(\ref{Ic}), by inserting Eq.(\ref{phicross}) into
Eq.(\ref{phisum}) and setting $\alpha=\pi/2$. The resulting
$I_c(H_z)$ is shown by the open squares in Fig.\ref{Crosstransverse}. Very similar results are obtained, if
Eq.(\ref{phicross}) is replaced by Eq.(\ref{Miller}). For large
fields, the calculations are well fitted by an inverse
proportionality law, $I_c \propto 1/H_z$, while the best fit to the
experimental data results in a quadratically decreasing dependence,
$I_c \propto H_z^{-2}$. Again we come to the conclusion that the
empirical expression taken from Ref.\cite{JAP08} is only valid in
the small field range, in this case, as far as $I_c(H_z)>0.2 I_0$.
Also for cross junctions the $I_c(H_a,\alpha)$ were calculated and
the $\alpha$ dependencies of $\Delta_R$ were extracted and shown by the 
open stars in Figs.\ref{CrossJunctions}. In evaluating the
integral (\ref{Ic}), the parameter $\eta$ in Eq.(\ref{phisum1}) was
chosen to be equal to its experimental counterpart
$\Delta_{R}^{\bot} /{\Delta_{R}^{\|}}$. As expected, the calculated
$\Delta_R(\alpha)$ are symmetric with respect to $\alpha=90^\circ$.

\subsection{Annular junctions\label{sec:AnnJun_T}}

\noindent In this section we will examine the static behavior of a small
annular JTJ in the presence of an oblique magnetic field.
Denoting the inner and outer ring radii, respectively, as $r_i$ and
$r_o$, we assume that the annular junction is unidimensional, i.e.,
the ring mean radius $\overline{r}= (r_i+r_o)/2$ is much larger than
the ring width $\triangle r = r_o-r_i$.

\noindent Using polar coordinates $r$ and $\theta$ such that $x=r
\cos \theta$ and $y= r \sin \theta$, the Josephson magnetic equation
(\ref{gra}) can be split into:

\begin{equation}
\label{Josann} {\frac{\partial \phi}{\partial r}} =\kappa H_\theta
\quad  \textrm{and} \,\,\,\, \quad
  {\frac{\partial \phi}{r \partial \theta}} =-\kappa H_r ,
\end{equation}

\noindent where $H_r$ and $H_\theta$ are the radial and tangential
components of the magnetic field in the ring plane, respectively.
With the annulus unidimensional, we can neglect the radial
dependence of the Josephson phase\cite{prb96}, i.e.,
$\phi(r,\theta)=\phi(\overline{r},\theta)$; henceforth,

\begin{equation}
\label{phiann} \phi(\theta)=-\kappa \overline{r} \int d\theta
H_r(\overline{r},\theta) + \phi_0 ,
\end{equation}

\noindent in which $\phi_0$ is an integration constant. By resorting
again to the superposition principle, we can readily write the
analogous of Eqs.(\ref{Hsum}) and (\ref{phisum}) for annular
junctions as:

\begin{equation}
\label{HsumA} {H_r}(\theta)={H_r^\|}(\theta)+{H_r^\bot}(\theta)
\end{equation}

\noindent and

\begin{equation}
\label{phisumA}  \phi(\theta)= \phi^\|(\theta) + \phi^\bot(\theta),
\end{equation}

\noindent provided $\partial \phi^\| /\partial \theta = -\kappa \overline{r} {H_r^\|}$
and $\partial \phi^\bot / \partial \theta = -\kappa \overline{r} {H_r^\bot}$.

\noindent It is well known\cite{prb96} that, when an external field is applied
in the plane of an electrically short ($\overline{r} < \lambda_J$)
annular junction, it fully penetrates the barrier, ${\bf H^\|}=H_x
{\bf \hat{x}}$, whose radial component ${H_r^\|}=H_x \cos\theta$, through (\ref{phiann}),
leads to:

\begin{equation} \label{phiparAnn}
\phi^\|(\theta) = h^\| \sin \theta,
\end{equation}

\noindent with $h^\| =\kappa H_x  \overline{r}$. As far as
$\phi(\theta)$ is an odd (periodic) function the calculation of the maximum critical current reduces to the following integration:

\begin{equation}
\label{annpatt} I_c=\frac{I_0}{\pi}\int_{0}^{\pi} d\theta \cos
\phi(\theta).
\end{equation}

\noindent Inserting $\phi$ as in (\ref{phiparAnn}), we obtain the
in-plane magnetic modulation pattern:

\begin{equation} \label{IcAnn}
I_c(h^\|)= I_0 \bigl| J_0(h^\|)\bigr|,
\end{equation}

\noindent  in which $J_0$ is the zero order Bessel function (of
first kind). In deriving (\ref{IcAnn}), it was assumed that the
Josephson current density is uniform over the barrier area and that
no magnetic flux is trapped in between the junction electrodes
[$\phi(\theta + 2 \pi)=\phi(\theta)$]. The static properties of an
annular junction in a transverse field have been numerically investigated in
Ref.\cite{JAP08} for a \textit{Lyngby} type annular JTJ obtained by
two films having the same widths\cite{davidson}; it was found that,
in a first approximation, $H_r^\bot$ sinusoidally depends on
$\theta$ resulting in a Bessel-like transverse magnetic pattern. In
other words, small differences are expected in comparing the shapes
of the in-plane and transverse magnetic diffraction patterns of an
annular junction, as shown by the logarithmic graph in Fig.\ref{annularComparison}. However, a small amplitude third
$\theta$-harmonic has to be added in order to correctly reproduce
$H_r^\bot(\theta)$:

\begin{equation} \label{Hradial}
H_r^\bot(\theta) \propto H_z (\cos \theta +  3 \delta
\cos 3\theta) ,
\end{equation}

\noindent with the coefficient $\delta$ much smaller than unity. The
last expression readily provides a more realistic
$\phi^\bot(\theta)$ dependence:

\begin{equation} \label{phiperpAnn}
\phi^\bot(\theta) =  h^\bot ( \sin \theta + \delta \sin 3\theta),
\end{equation}

\noindent with $h^\bot \propto H_z$. Being $H_x=H_a \cos \alpha$ and $H_z=H_a \sin \alpha$,
by inserting (\ref{phiparAnn}) and (\ref{phiperpAnn}) into
(\ref{phisumA}), the Josephson phase in the presence of an arbitrary oblique magnetic field applied with amplitude $H_a=\sqrt{H_x^2+H_z^2}$ and orientation $\alpha= \arctan
(H_x/H_z)$ becomes:

\begin{equation}
\label{phisumAnn}  \phi_{h,\alpha}(\theta)= h \left[ \sin\theta \,
\cos\alpha + \eta (\sin \theta + \delta \sin 3\theta)
\sin\alpha\right],
\end{equation}

\noindent in which, again, $h^\|$ has simply been renamed $h$
and $\eta= h^\bot/h^\|$. In order to reproduce the experimental
findings reported in Sec.\ref{sec:AnnJun}, we have computed the
$I_c(h,\alpha)$, inserting Eq.(\ref{phisumAnn}) in
Eq.(\ref{annpatt}), being still $ \phi_{h,\alpha}(-\theta)=
-\phi_{h,\alpha}(\theta)$. The value of $\eta$ has been taken from
the experimental $\Delta_{A}^{\bot}/\Delta_{A}^{\|}$ ratio:
$\eta_A=0.63$ from Table I, while the value of $\delta$ was
determined from the best fit of the experimental transverse magnetic pattern,
$\delta=0.02$. The results of such calculations for several
$\alpha$ values are displayed in Fig.\ref{annular} with open
stars; the agreement with the experimental data is excellent.
Indeed, the third harmonic correction is mainly needed to reproduce
the peak in $\Delta_A(\alpha)$; in fact, with $\delta$ set to zero,
the field amplitude dependence in Eq.(\ref{phisumAnn}) would be
undetermined for $\alpha= -\arctan {1}/{\eta}$, resulting in an
unphysical independence of $I_c$ on $H_a$. However, far enough from
this critical angle, we can set $\delta=0$ in Eq.(\ref{phisumAnn}) and
the integral (\ref{annpatt}) trivially results in a Bessel-like
behavior with critical field $H_c$ or pattern width $\Delta_A$ given
by:

\begin{equation}
\label{Deltalfa} {\Delta_A(\alpha)} = \frac {\Delta_{A}^{\|}}
{\left| \cos{\alpha} + \eta \sin{\alpha} \right|} .
\end{equation}

\noindent The last approximate expression, plotted as a the solid
line in Fig.\ref{annular}, exactly matches the experimental
points, everywhere except near $\alpha_M =-\arctan (1/\eta_A) \simeq
145^\circ$, where it goes to infinity. From measurements not
reported in this paper, we found that Eq.(\ref{Deltalfa}) can
be usefully applied to reproduce also the behavior of long ($\overline{r}
> 10 \lambda_J$) unidimensional Lyngby-type annular junctions in
the presence of an arbitrary oblique field.

\section{Discussion\label{sec:Discussion}}

\noindent A similar approach can be adopted to describe the behavior
of overlap-type junctions having the aspect ratio smaller than
unity. In fact, being $-1\leq \bar{x} \leq 1$ and $-1 \leq \bar{y}
\leq 1$, then $\sin\bar{y} \approx \bar{y}$, and, under the
assumption $\beta<1$, $\cosh\beta \bar{x}\approx 1$. Therefore,
Eq.(\ref{phisum1}) simplifies to a linear $\bar{y}$ dependence:

$$ \phi(\bar{x},\bar{y})= h( \cos\alpha + \eta \sin\alpha) \bar{y}, $$

\noindent resulting in a Fraunhofer-like magnetic diffraction
pattern with $H_c$ or, with our notation, $\Delta_R$, given by:

\begin{equation}
\label{Deltavsalfa} {\Delta_R(\alpha)} = \frac {\Delta_{R}^{\|}}
{\left| \cos{\alpha} + \eta \sin{\alpha} \right|} =  {1}/
{\left| \frac{\cos\alpha}{\Delta_{R}^{\|}} + \frac{
\sin\alpha}{\Delta_{R}^{\bot}} \right|}.
\end{equation}

\noindent It's seen that the critical width $\Delta_R$ diverges at a
critical angle $\alpha_c$ given by:

\begin{equation}
\label{alfac}
\alpha_c = -\arctan \frac{1}{\eta} =-\arctan\frac{\Delta_R^{\|}}{\Delta_R^\bot}.
\end{equation}

\noindent In other words, when $\alpha=\alpha_c$, the effect of the
transverse component of the applied magnetic field exactly cancels
that of the in-plane component, so that the junction is virtually
insensitive to the magnetic field. In practice, a 
full compensation is never achieved, although $\Delta_R(\alpha_M)$
can be as large as $100 \Delta_{R}^{\|}$. Eq.(\ref{Deltavsalfa}) is
plotted as a solid line in Figs.\ref{OverDeltaAlfa}a-c and it is
seen that it nicely reproduces the experimental data, not only when
$\beta<1$ (see Fig.\ref{OverDeltaAlfa}b), but also for $\beta=1$
(see Fig.\ref{OverDeltaAlfa}a). It also closely fits the results
of a long ($2W \simeq 12 \lambda_J$) unidimensional JTJ (see
Fig.\ref{OverDeltaAlfa}c) for which the calculations developed
in Sec.\ref{sec:OverJun_T} for electrically small junctions do not
apply. Eqs.(\ref{Deltalfa}) and (\ref{Deltavsalfa}) indicate that,
for small unidimensional annular junctions and small overlap type
junctions with small aspect ratio, only two measurements are needed
to forecast their static behavior in an arbitrary oblique field, that is,
the in-plane and the transverse magnetic diffraction patterns.

As far as cross-type samples $JJ\#E$ and $JJ\#F$ are concerned, we
observe that the experimental data shown, respectively, in
Fig.\ref{OverDeltaAlfa}(a) and (b) are affected by a slight
asymmetry with respect to $\alpha=90^\circ$, in contrast with the
system symmetry properties (even in the case of electrically long
junctions). We explain this symmetry break in terms of tiny
fabrication misalignments. In our fabrication line, once the base
electrode has been etched away, each next layer is positioned with
an accuracy better than $1\, \mu m$ (in each direction). This
accuracy, although smaller, is comparable with the idle-region
dimension, resulting in an unavoidable, albeit small, imperfection
in positioning the junction area exactly in the centers of the
films. Since the screening currents mainly flow along the electrode
borders, the misalignment effect increases with the junction side
dimension. This explains why the observed asymmetry is larger for
junction $\#F$.

For the Lyngby-type annular junction our theory reproduces the
experimental data in a more than satisfactory fashion. However, the
situation might be not so good for annular junctions made by
electrodes of unequal widths as, for example, those used in
Ref.\cite{PRL06}, which require also the introduction of the second
$\theta$-harmonic in Eq.(\ref{Hradial}). However, when 
$\phi^\bot(\theta)$ is as in (\ref{phiperpAnn}), exploiting the
trigonometric equivalence $\sin 3x =\sin x (1+2\cos 2x) $, the
transverse magnetic pattern can be analytically shown to be given by
the following even expression:

$$ I_c(h^\|)= I_0 \bigl| J_0(h^\|) + 2 \delta J_2(h^\|) \bigr|, $$

\noindent in which $J_2$ is the second-order Bessel function and
$\delta$ was assumed to be much smaller than unity. Similarly,
$I_c(h)$ can be analytically worked out in the more general case of
Eq.(\ref{phisumAnn}), as far as $\alpha$ is far from $\alpha_M$,
more precisely, when $\delta << 1+1/ \eta_A \tan \alpha$.

The weak point of our theoretical approach based on the
superposition principle is the indetermination of the parameter
$h^\bot$ introduced in Eq.(\ref{phiperp}). The knowledge of the
magnetic field actually introduced into the barrier for a given
transverse field $H_{z}$ requires a careful experimental
investigation with samples having a given junction geometry and
different geometrical parameter of the connecting electrodes.
However, while the superconducting film widths can be easily varied,
it is difficult to realize samples with much different film
thicknesses. Consequentially, when the shape of the $I_c(H_z)$ is
known \textit{a priori}, the effective normalized field $h^\bot$
remains to be determined from the direct measure of the transverse
magnetic diffraction pattern.

\section{Conclusion\label{sec:Conc}}

In this paper we examined the static properties of small planar
Josephson tunnel junction in presence of a uniform external field
applied at an arbitrary angle with respect to the barrier plane.
This topic has been considered from both the experimental and
theoretical point of view. We have presented the recorded oblique
magnetic diffraction patterns of junctions with the most common
electrode configuration, namely overlap, cross and annular
geometries. These data, beside being original by themselves, also
served as a test for the theoretical analysis of small JTJs in a
purely transverse field that we recently proposed\cite{JAP08}.  Further, by invoking the superposition principle, the findings of
Ref.\cite{JAP08} in a transverse field were combined with the
classical knowledge for a JTJ in a parallel field to provide a
general theory for any arbitrary oblique field. We stress that the
theory has been developed assuming that junction was electrically
small, naked, geometrically perfect and made with purely diamagnetic
superconducting films ($\lambda_L=0$ and $H_c=\infty$). Although our
samples satisfied these conditions only to a rough approximation,
the agreement between the experimental and theoretical results is
more than satisfactory especially for annular junctions. For
rectangular samples the theoretical approach is again very good only
for small values of the applied field and of the junction aspect
ratio. In the other cases, our theoretical predictions fail to
provide a correct description. This was to be expected because the
approximate analytical expressions heuristically found in
Ref.\cite{JAP08} were already observed to have the largest relative
error near the junction corners. As already reported
elsewhere\cite{Miller85}, the importance of the Josephson phase at the
junction corners grows with the amplitude of the external field.

\noindent The main message of this paper is that even a small
transverse field (which has been largely ignored in the past) can
strongly influence the magnetic interference patterns. We explore
the implications of this result in supposing systematic errors in
previous experiments and in proposing new possible applications. Since in most of the applications the external field
needed to modulate the critical current of a planar JTJ is applied
in the barrier plane by means of a long solenoid or Helmholtz coil
pairs, we want to stress the importance of the alignment of the
solenoid (or coil) axis with the junction plane. It was believed
that any possible tiny angular misalignment $\delta\alpha$ between
the coil axis and the junction plane would result in only
second-order errors, being $H_{x} = (1 - 0.5
{\delta\alpha}^2)H_{a}$. However, as we have shown, this is not
necessarily true, since the unwanted (and often unknown) small
transverse field component $H_{z} = \delta\alpha H_{a}$ might have
an effect comparable or even larger than that of the in-plane
component, if the junction critical angle $\alpha_c$ is close to
$180^\circ$. For example, in the case of sample $JJ\#C$ of Table I,
a misalignment $\delta\alpha \simeq -4^\circ$ would result in a
systematic error of more than one order of magnitude for the
measurement of the first critical field. The consequence of the coil
misalignment might have been underestimated, if not ignored, in many
previous experiments dealing with JTJs in an external magnetic
field, including those in which the exact knowledge of the field in
the barrier plane is of capital importance. As a corollary, it also
follows that in shielding a cryoprobe the same care has to be taken
to minimize both the in-plane and the transverse stray fields.

\noindent Furthermore, in planar SQUID applications the magnetic
field to be measured is applied perpendicular to the SQUID loop and
its effect on the junction critical currents has never been
considered. However, it might not be negligible, especially when
the field is large and the junction(s) is placed close to the
borders of the superconducting electrodes where the induced
screening currents are larger - this is the case of step-edge or
ramp-type junctions\cite{podt,smilde}.

In this paper we have shown that for a given junction
geometry the response to an externally applied magnetic field
drastically depends on the field orientation, a property that might
be exploited to design angle resolving instruments. Furthermore,
considering that this response is different for JTJs having
different geometries it makes possible to design multijunction chips
in which, for a given applied magnetic field, the critical current of some
junctions is almost completely suppressed while that of other
junctions remains unaffected. Our findings also suggest that two (or
more) independent magnetic fields with different amplitudes and
orientations can be applied to multijunction chips in order to
obtained the proper critical current suppression required for
samples having different geometrical configurations.

\noindent By using Eq.(\ref{Paterno}), the theory developed in
Sec.\ref{sec:Theory} for a uniform oblique field applied in the
$y=0$ plane, can be easily extended to the most general case in
which all three field components are non-zero. In other words, the
oblique magnetic diffraction pattern can be theoretically predicted
for the most common junction configurations, as far as the junction
dimensions are smaller than Josephson penetration length.  The case
of long JTJ in a transverse field still remains an open question
since it requires the solution of a tridimensional magnetostatic
problem in the presence of external (non-linear) currents\cite{goldobin}.

\section*{Acknowledgements}

We thank Pavel Dmitriev for the fabrication of the samples.

\section*{Note}

Soon after this paper was accepted for publication in Phys. Rev. B we become aware of a paper by 
Heinsohn \textit{et al.} entitled "Effect of the magnetic-field orientation on the modulation period of the critical current of ramp-type Josephson junctions " (see Ref.27). The authors report several experimental magnetic diffraction patterns of both low and high temperature ramp-type Josephson tunnel junctions. As first found by Rosestein and Chen, they observed that the pattern modulation period drastically changes with the external field orientation.

\newpage

\end{document}